# Regression approaches for modelling genotype-environment interaction and making predictions into unseen environments

**Maksym Hrachov[1], Hans-Peter Piepho[1], Niaz Md. Farhat Rahman[2], Waqas Ahmed Malik[1]**
1 Biostatistics Unit, Institute of Crop Science, University of Hohenheim, Stuttgart, Germany
2 Bangladesh Rice Research Institute (BRRI), Gazipur, Bangladesh

**ORCID**
Maksym Hrachov: 0009-0002-5410-5732
Hans-Peter Piepho: 0000-0001-7813-2992
Niaz Md. Farhat Rahman: 0000-0002-8789-4144
Waqas Ahmed Malik: 0000-0001-6455-5353

**Correspondence**
Hans-Peter Piepho, Biostatistics Unit, Institute of Crop Science, University of Hohenheim, 70593 Stuttgart, Germany.
Email: piepho@uni-hohenheim.de

**Author contribution statement**
HPP reviewed and compared regression methods and developed methods for computing individual and pairwise prediction variances.
MH implemented all models and methods in statistical software, conducted cross-validation, summarized the results, prepared appendices and example code.
NMFR and WAM compiled the dataset from historical rice trial records and assisted in acquiring environmental covariates for the trial locations.

**Abstract**

In plant breeding and variety testing, there is an increasing interest in making use of environmental information to enhance predictions for new environments. Here, we will review linear mixed models that have been proposed for this purpose. The emphasis will be on predictions and on methods to assess the uncertainty of predictions for new environments. Our point of departure is straight-line regression, which may be extended to multiple environmental covariates and genotype-specific responses. When observable environmental covariates are used, this is also known as factorial regression. Early work along these lines can be traced back to Stringfield & Salter (1934) and Yates & Cochran (1938), who proposed a method nowadays best known as Finlay-Wilkinson regression. This method, in turn, has close ties with regression on latent environmental covariates and factor-analytic variance-covariance structures for genotype-environment interaction. Extensions of these approaches – reduced rank regression, kernel- or kinship-based approaches, random coefficient regression, and extended Finlay-Wilkinson regression – will be the focus of this paper. Our objective is to demonstrate how seemingly disparate methods are very closely linked and fall within a common model-based prediction framework. The framework considers environments as random throughout, with genotypes also modelled as random in most cases. We will discuss options for assessing uncertainty of predictions, including cross validation and model-based estimates of uncertainty, the latter one being estimated using our new suggested approach. The methods are illustrated using a long-term rice variety trial dataset from Bangladesh.

**Key message:** several seemingly distinct regression methods are closely related. Environmental covariates delivered improved prediction, and a new approach improves estimation of prediction variance.

## 1 Introduction

In plant breeding and variety testing, there is an increasing interest in making use of environmental information to enhance prediction of varietal performance. Key words associated with methods to implement this predictive objective are *envirotyping* (Xu 2016) and *enviromics* (Cooper & Messina 2021; Resende et al. 2021). The purpose of this paper is to show how different models for genotype-environment interaction that make use of environmental covariate (EC) information for prediction into a target population of environments (TPE) are related. While we focus on a set of linear mixed models, other linear as well as machine and deep learning approaches also exist in this field (Hu et al. 2025; Yu et al. 2025; Zou et al. 2025). The models we discuss in this paper can readily accommodate nonlinearities by including quadratic terms for the regressors (Buntaran et al. 2021). Interactions among regressors can also be incorporated, leading to a response surface regression framework (Box & Draper 2007).
The point of departure will be the factorial regression (FR) model (Denis et al. 1997). Subsequently, modelling genotype as a random factor, three different variance-covariance structures can be imposed, leading to three different models, which are denoted here as *random FR (RFR)*, *environmental kernel approach* (Jarquín et al. 2014) and *reduced rank regression* (RRR) (Buntaran et al. 2021; Tolhurst et al. 2022). We also make a link between RRR and a recent paper (Piepho & Blancon 2023) that proposed an extension of Finlay-Wilkinson regression (Finlay & Wilkinson 1963; Yates & Cochran 1938). Furthermore, we consider the variance of predictions using EC under four basic scenarios. The framework is illustrated using data from a multi-environment trial (MET) in Bangladesh.

## 2 Regression models

### 2.1 Factorial regression with random genotypes as the point of departure

FR model involves a genotype-specific multiple linear regression on EC (Denis et al. 1997; Piepho & Blancon 2023), and can be written as

$$\eta_{ij} = \alpha_i + \gamma_{i1}x_{j1} + \gamma_{i2}x_{j2} + \ldots + \gamma_{ip}x_{jp} \ , \tag{1}$$

where $\eta_{ij}$ is the expected response for the $i$-th genotype $(i = 1, \ldots, I)$ in the $j$-th environment $(j = 1, \ldots, J)$, $\alpha_i$ is the intercept for the $i$-th genotype, $\gamma_{ik}$ is the slope for the $k$-th environmental covariate (EC) for the $i$-th genotype, and $x_{jk}$ is the value of the $k$-th covariate $(k = 1, \ldots, p)$ for the $j$-th environment.

The FR model in eq. (1) states the expected response $\eta_{ij}$ for given values of the EC and does not include a deviation from the regression. Such deviations are certainly needed to model the observed response, which may be denoted as $y_{ij}$. For making predictions into other environments in a defined TPE, and at the same time assessing the uncertainty of predictions based on a fitted mixed model, it is necessary to model environments as random (Piepho & Williams 2024). This approach is in spirit related to small area estimation techniques detailed in a book by Rao & Molina (2015) which includes an extensive treatment of model-based mean squared error estimation. If environments are random, deviations from the regression are also random. Moreover, deviations of different genotypes from the regression in the same environment are likely to be positively correlated. The simplest way to account for such correlation is to add a random environmental main effect $u_j$. The full model for observed data then becomes

$$y_{ij} = \eta_{ij} + u_j + e_{ij} \ , \tag{2}$$

where $y_{ij}$ is the observed mean for the $i$-th genotype in the $j$-th environment and $e_{ij}$ is a random residual. The two random effects jointly model the deviation from the regression lines, given by $d_{ij} = u_j + e_{ij}$ (Piepho & Blancon 2023). Note in passing that environments are usually indexed by both years and locations. If this is the case, it is useful to employ a full factorial model for $u_j$ and $e_{ij}$, which, when taking into consideration the factor genotype for the full model, leads to a three-way model (Talbot 1984; Piepho & Williams 2024). This factorization will be considered in Section 3.5.

Regarding genotypes, one option is to model them as fixed. This kind of model with fixed genotypes and random environments was considered by Denis et al. (1997) and Piepho et al. (1998). Alternatively, genotype may be modelled as a random factor. This model can be re-written as

$$\alpha_i = \mu_\alpha + a_i \ , \tag{3}$$
$$\gamma_{ik} = \mu_{\gamma k} + c_{ik} \ , \text{ and} \tag{4}$$
$$\eta_{ij} = (\mu_\alpha + a_i) + (\mu_{\gamma 1} + c_{i1})x_{j1} + (\mu_{\gamma 2} + c_{i2})x_{j2} + \ldots + (\mu_{\gamma p} + c_{ip})x_{jp} \ , \tag{5}$$

where

$$\begin{pmatrix} a_i \\ \boldsymbol{c}_i \end{pmatrix} = N\left[\begin{pmatrix} 0 \\ \mathbf{0}_p \end{pmatrix}, \boldsymbol{\Sigma}\right], \tag{6}$$

where $\boldsymbol{c}_i = (c_{i1}, \ldots, c_{ip})^T$, super-scripted $T$ denotes the transpose of a vector or matrix, and $\mathbf{0}_p$ is a $p$-vector of zeros, and $\boldsymbol{\Sigma}$ is a variance-covariance matrix (common to all genotypes). Rearranging terms and defining $\boldsymbol{x}_j = (x_{j1}, \ldots, x_{jp})^T$, this model can be written as

$$\eta_{ij} = f(\boldsymbol{x}_j) + g_i(\boldsymbol{x}_j)\,, \tag{7}$$

where the fixed-effects part

$$f(\boldsymbol{x}_j) = \mu_\alpha + \mu_{\gamma 1} x_{j1} + \mu_{\gamma 2} x_{j2} + \ldots + \mu_{\gamma p} x_{jp} \tag{8}$$

corresponds to a mean regression across genotypes, whereas the random part

$$g_i(\boldsymbol{x}_j) = a_i + c_{i1} x_{j1} + c_{i2} x_{j2} + \ldots + c_{ip} x_{jp} \tag{9}$$

models the random deviation of the regression for the $i$-th genotype from the mean regression $f(\boldsymbol{x}_j)$. An important observation at this point is that the model has a random main effect for genotypes ($a_i$). If we include the random environmental main effect $u_j$ for the observed response, the model for the response $y_{ij}$ has a mixed main effect for environments, given by

$$\varepsilon_j = \mu_{\gamma 1} x_{j1} + \mu_{\gamma 2} x_{j2} + \ldots + \mu_{\gamma p} x_{jp} + u_j\,, \tag{10}$$

where the random effect $u_j$ acts as a random deviation from the mean regression $f(\boldsymbol{x}_j)$ in eq. (8).

### 2.2 Specification of the variance-covariance structure for $\eta_{ij}$

If we collect responses $\eta_{ij}$ for the $i$-th genotype into a vector $\boldsymbol{\eta}_i = (\eta_{i1}, \eta_{i2}, \ldots, \eta_{iJ})^T$ ordered by environments, we have

$$\boldsymbol{\eta}_i = \mathbf{1}_J(\mu_\alpha + a_i) + \mathbf{X}(\boldsymbol{\mu}_\gamma + \boldsymbol{c}_i), \text{ and} \tag{11}$$

$$\boldsymbol{\eta}_i \sim N\left[\mathbf{1}_J \mu_\alpha + \mathbf{X}\boldsymbol{\mu}_\gamma, \boldsymbol{\Omega}\right], \tag{12}$$

where $J$ is the number of environments, $\mathbf{1}_J$ is a $J$-vector of ones,

$$\mathbf{X} = \{x_{jk}\} = \begin{pmatrix} x_{11} & x_{12} & \cdots & x_{1p} \\ x_{21} & x_{22} & \cdots & \vdots \\ \vdots & \vdots & \ddots & \vdots \\ x_{J1} & x_{J2} & \cdots & x_{Jp} \end{pmatrix}, \tag{13}$$

$\boldsymbol{\mu}_\gamma = (\mu_{\gamma 1}, \mu_{\gamma 2}, \ldots, \mu_{\gamma p})^T$, and

$$\boldsymbol{\Omega} = (\mathbf{1}_J \vdots \mathbf{X})\boldsymbol{\Sigma}(\mathbf{1}_J \vdots \mathbf{X})^T. \tag{14}$$

Note that $(\mathbf{1}_J \vdots \mathbf{X})$ is column-wise concatenation of $\mathbf{1}_J$ and $\mathbf{X}$. We will assume here that all $p$ EC in $\mathbf{X}$ have been mean-centered and scaled to unit variance. There are different possible specifications for the $(p+1) \times (p+1)$ variance-covariance matrix $\boldsymbol{\Sigma}$ in (6), and consequently different forms of $\boldsymbol{\Omega}$.

(i) In random coefficients regression, $\boldsymbol{\Sigma}$ is chosen to be unstructured in order to ensure translational invariance, i.e. invariance to linear transformations of the covariates (Buntaran et al. 2021; Longford 1993). We denote this approach in our context as RFR.

(ii) By comparison, the *kernel approach* (Jarquín et al. 2014) assumes

$$\boldsymbol{\Sigma} = \begin{pmatrix} \sigma_\alpha^2 & 0 \\ 0 & \mathbf{I}_p \sigma_\gamma^2 \end{pmatrix}. \tag{15}$$

This model clearly is not translationally invariant, i.e. linear transformations of the columns in $\mathbf{X}$ would alter the fit. However, it only has two parameters and is therefore more parsimonious than the RFR model with unstructured $\boldsymbol{\Sigma}$. In fact, the kernel approach can be regarded as the most simplistic reduction of the RFR model.

The kernel model can also be motivated by a regularization argument and implies a ridge regression (Ruppert et al. 2003, p.66) on the EC. We further find that

$$\boldsymbol{\Omega} = \mathbf{1}_J\mathbf{1}_J^T\sigma_\alpha^2 + \mathbf{X}\mathbf{X}^T\sigma_\gamma^2 \ . \tag{16}$$

The matrix $\mathbf{K}_E = \mathbf{X}\mathbf{X}^T$ is seen to be the kernel matrix for environments. Note that standardizing the columns of $\mathbf{X}$ to zero mean and unit variance makes the kernel approach unique and ensures that each EC has equal influence on the regression, but does not resolve the lack of translational invariance issue. The model can be re-written in scalar form as

$$\eta_{ij} = \mu_\alpha + \mu_{\gamma 1}x_{j1} + \mu_{\gamma 2}x_{j2} + \ldots + \mu_{\gamma p}x_{jp} + a_i + w_{ij}, \tag{17}$$

where $a_i \sim N\,(0, \sigma_\alpha^2)$ is a genotype main effect and $\boldsymbol{w}_i = \left(w_{i1}, w_{i2}, \ldots, w_{iJ}\right)^T \sim N\left(0, \mathbf{K}_E\sigma_\gamma^2\right)$ is the vector of interactions for the $i$-th genotype. Note that the model in (17) involves the mean regression $f\left(\boldsymbol{x}_j\right)$. Authors applying the kernel approach often omit this mean regression, implying that the expected value of the regression coefficients over genotypes is zero. This assumption is usually unrealistic. A non-zero expectation can be incorporated into the model either explicitly, by including the fixed mean regression for environmental covariates as described here, or implicitly, by including fixed main effects for environments. It must be borne in mind, however, that fitting the mean regression $f\left(\boldsymbol{x}_j\right)$ requires that the number of environments exceeds the number of covariates. The environmental kernel approach is most often applied in cases where this condition does not hold.

(iii) There is an intermediate option between RFR and the kernel approach, which may be termed RRR. For it we may use

$$\boldsymbol{\Sigma} = \boldsymbol{\Lambda}\boldsymbol{\Lambda}^T, \tag{18}$$

where $\boldsymbol{\Lambda}$ is a $(p+1) \times q$ matrix of factor loadings for $q$ latent factors and $(p+1)$ regression terms (one intercept, $p$ slopes). The RRR model can also be denoted as a factor-analytic (FA) model with no residual variances. Essentially, this model approximates the unstructured model for $\boldsymbol{\Sigma}$ in RFR using a reduced rank matrix (Buntaran et al. 2021; Tolhurst et al. 2022). It is worth stressing that the RRR approximation to $\boldsymbol{\Sigma}$ is also translationally invariant (Tolhurst 2023). Figure A1 in Appendix A illustrates the connection between these three models.

**Two-stage approach to fit the RRR:** If we use the partition

$$\boldsymbol{\Lambda} = \begin{pmatrix} \boldsymbol{\lambda}_\alpha^T \\ \boldsymbol{\Lambda}_\gamma \end{pmatrix} = \begin{pmatrix} 1 & 0 \\ 0 & \boldsymbol{\Lambda}_\gamma \end{pmatrix} \begin{pmatrix} \boldsymbol{\lambda}_\alpha^T \\ \mathbf{I}_q \end{pmatrix} \ , \tag{19}$$

where $\boldsymbol{\lambda}_\alpha = \left(\lambda_{\alpha(1)}, \ldots, \lambda_{\alpha(q)}\right)^T$ is a $q$-vector of loadings pertaining to the intercept and $\boldsymbol{\Lambda}_\gamma = \left\{\lambda_{\gamma(kh)}\right\}$ $(h = 1, \ldots, q)$ is the $p \times q$ sub-matrix of $\boldsymbol{\Lambda}$ pertaining to the slopes, then we may represent the model for the random effects $a_i$ and $\boldsymbol{c}_i$ by

$$\begin{pmatrix} a_i \\ \boldsymbol{c}_i \end{pmatrix} = \boldsymbol{\Lambda}\boldsymbol{v}_i, \tag{20}$$

where $\boldsymbol{v}_i \sim N\left(\mathbf{0}_q, \mathbf{I}_q\right)$. Note that this representation implies the reduced rank structure for $\boldsymbol{\Sigma}$ in (18). We find after some re-arrangement that

$$a_i = \boldsymbol{\lambda}_\alpha^T\boldsymbol{v}_i \ , \tag{21}$$
$$\boldsymbol{c}_i = \boldsymbol{\Lambda}_\gamma\boldsymbol{v}_i, \tag{22}$$
$$\mathbf{X}\boldsymbol{c}_i = \mathbf{X}\boldsymbol{\Lambda}_\gamma\boldsymbol{v}_i = \mathbf{Z}\boldsymbol{v}_i \tag{23}$$

where $\mathbf{Z} = \left\{z_{jh}\right\} = \mathbf{X}\boldsymbol{\Lambda}_\gamma$ with

$$z_{jh} = \lambda_{\gamma(1h)}x_{j1} + \lambda_{\gamma(2h)}x_{j2} + \ldots + \lambda_{\gamma(ph)}x_{jp} \tag{24}$$

the value of the $h$-th synthetic environmental covariate (SC) for the $j$-th environment (Piepho & Blancon 2023), and hence

$$\boldsymbol{\eta}_i = \mathbf{1}_J \mu_\alpha + \mathbf{X}\boldsymbol{\mu}_\gamma + \mathbf{1}_J a_i + \mathbf{Z}\boldsymbol{v}_i \ . \tag{25}$$

Also note that

$$var\begin{pmatrix} a_i \\ \boldsymbol{v}_i \end{pmatrix} = \begin{pmatrix} \boldsymbol{\lambda}_\alpha^T \boldsymbol{\lambda}_\alpha & \boldsymbol{\lambda}_\alpha^T \\ \boldsymbol{\lambda}_\alpha & \mathbf{I}_q \end{pmatrix} = \widetilde{\boldsymbol{\Lambda}}\widetilde{\boldsymbol{\Lambda}}^T \tag{26}$$

with $\widetilde{\boldsymbol{\Lambda}}^T = (\boldsymbol{\lambda}_\alpha \quad \mathbf{I}_q)$ and

$$\boldsymbol{\Omega} = \left(\mathbf{1}_J \vdots \mathbf{Z}\right)\widetilde{\boldsymbol{\Lambda}}\widetilde{\boldsymbol{\Lambda}}^T\left(\mathbf{1}_J \vdots \mathbf{Z}\right)^T , \tag{27}$$

This is recognized as a RRR using the synthetic covariates $\mathbf{Z}$. The important practical implication of this result is that we can use the approach described in Piepho & Blancon (2023) to derive the synthetic covariates, which involves estimating $\boldsymbol{\Lambda}_\gamma$ and then fit (27), which involves estimating $\boldsymbol{\lambda}_\alpha$. This can then be regarded as a two-stage approach to fit the RRR. To exploit this with a mixed model package, it is necessary to keep in mind how the package imposes constraints on $\widetilde{\boldsymbol{\Lambda}}^T$. For example, In ASReml-R and SAS, one needs to permute the synthetic covariates $\mathbf{Z}$ and the intercept $\mathbf{1}_J$ to be able to fit the model at the second stage.

**Extended Finlay-Wilkinson regression**: Piepho and Blancon (2023) suggested to obtain the synthetic covariates in (24) using (11) with $a_i$ taken as fixed and assuming $var(\boldsymbol{c}_i) = \boldsymbol{\Lambda}_\gamma\boldsymbol{\Lambda}_\gamma^T$. In addition, instead of fitting the mixed-effects mean regression according to eq. (10), one may consider fitting a simple fixed environmental main effect $\varepsilon_j$ to make sure the coefficients in $\boldsymbol{\Lambda}_\gamma$ are optimized to explain the genotype-environment interaction. From the fitted matrix $\boldsymbol{\Lambda}_\gamma$ one can then compute the SC using $\mathbf{Z} = \mathbf{X}\boldsymbol{\Lambda}_\gamma$. Subsequently, the model

$$\boldsymbol{\eta}_i = \mathbf{1}_J \alpha_i + \mathbf{Z}\boldsymbol{\beta}_i, \tag{28}$$

can be fitted, where $\alpha_i$ is a fixed intercept and $\boldsymbol{\beta}_i = \left(\beta_{i(1)}, \dots, \beta_{i(q)}\right)^T$ are fixed regression coefficients for the $i$-th genotype pertaining to the $q$ SC in $\mathbf{Z}$, assuming that genotype is a fixed factor and environments are random. Piepho and Blancon (2023) referred to this approach as *extended Finlay-Wilkinson (FW) regression*. We note that for implementing this approach, it is convenient to impose constraints on $\boldsymbol{\Lambda}$ via $\boldsymbol{\Lambda}_\gamma = \{\lambda_{kh}\}$, requiring that $\lambda_{kh} = 0$ for $k > h$. At the same time no constraints are imposed on $\boldsymbol{\lambda}_\alpha$ because $\alpha_i$ is modelled as fixed. Also note that the regression coefficients $\boldsymbol{\beta}_i$ in (28) are related to the coefficients $\boldsymbol{v}_i$ in (25).

**Adding genetic relationship matrix:** The models introduced so far assume independence between genotypes. This can be modified by assuming a kinship matrix $\mathbf{K}_G$ for genotypes. For example, under the environmental kernel approach for EC, the vector of interaction effects $\boldsymbol{w} = (w_1^T, w_2^T, \dots, w_I^T)^T$, where $I$ is the number of genotypes and $w_i$ is defined with the model in eq. (17), can be assumed to be distributed as

$$\boldsymbol{w} \sim N\left(\mathbf{0}_{IJ}, \mathbf{K}_G \otimes \mathbf{K}_E \sigma_\gamma^2\right) \ . \tag{29}$$

Similarly, for the genotype main effect $\boldsymbol{a} = (a_1, a_2, \dots, a_I)^T$ we may assume

$$\boldsymbol{a} \sim N\left(\mathbf{0}_I, \mathbf{K}_G \sigma_\alpha^2\right) \ . \tag{30}$$

The RFR and RRR approaches can be similarly modified.

## 3 Using regression models for predictions into new environments

### 3.1 A single regression term

One important use of regression models involving EC is to make predictions of genotype performances into new environments. The new environment may involve a new location, a new year, or both. Clearly, for giving out recommendations to farmers, the farm's location is almost invariably an unseen location because it is not part of the trial network based on which the recommendation is made. Similarly, the most relevant scenario is for the new environment to be a projection into a future year or set of years. We assume here that long-term data are available for all EC in all locations of the TPE. Assuming that environment is a random factor and given that environments are indexed by locations and years, a two-way model may be assumed for each EC. For simplicity, we here consider a single EC. Extension to multiple EC and also to multiple SC is straightforward as will be shown later. For a single EC, the model can be written

$$x_{lm} = \mu_x + L_{x(l)} + Y_{x(m)} + (LY)_{x(lm)} \ , \tag{31}$$

where $x_{lm}$ is the value of the EC in the $l$-th location in the $m$-th year, $\mu_x$ is an intercept, $L_{x(l)} \sim N\left(0, \sigma^2_{x(L)}\right)$ is the random main effect for the $l$-th location, $Y_{x(m)} \sim N\left(0, \sigma^2_{x(Y)}\right)$ is the random main effect for the $m$-th year, and $(LY)_{x(lm)} \sim N\left(0, \sigma^2_{x(LY)}\right)$ is the random location-year interaction for the $l$-th location and $m$-th year.

When making predictions of $\eta_i$, we need to consider that, depending on the prediction scenario, the exact value of $x$ may not be known with certainty for the targeted environment, but only a long-term mean of $x$, either for a specific location from the TPE that is the target of prediction, of for the whole of the TPE. Also, it may be possible that we are specifically interested in the prediction at a mean value of $x$, e.g. for a specific location or for the whole TPE. Alternatively, a prediction may be needed for a future year, in which case uncertainty regarding the future value of $x$ may come into play. In what follows, four different scenarios (cases) are distinguished. To set the stage, we will initially consider the regression term $\gamma_i x$ for a single EC and its contribution to the prediction, as well as to the uncertainty of the prediction. Extension to the full model, including the intercept, and multiple EC will be considered later.

In general, we will need to use a value to plug in for $x$. This value will be an expected value of $x$ and will be denoted as $\xi$, so in general the term $\gamma_i \xi$ is used to predict the mean performance. There may also be uncertainty associated with the value of $x$, and this can be assessed via a variance of $x$, which we will denote as $\sigma^2_x$, and which can be partitioned as $\sigma^2_x \ = \sigma^2_{x(L)} + \ \sigma^2_{x(Y)} + \sigma^2_{x(LY)}$. To assess the contribution $\phi_{x(i)}$ to the overall prediction variance (considered in detail in Section 3.2), we will initially assume that both $\gamma_i$ and $\xi$ are known, and hence the contribution to the prediction variance for the response is $\phi_{x(i)} = \gamma_i^2 \sigma^2_x$. In subsequent sections, we will account the additional uncertainty arising from the fact that both $\gamma_i$ and $\xi$ in the product $\gamma_i \xi$ need to be estimated.

In what follows, we will consider four different cases as regards the target of prediction. The four cases are summarized in Table 1. Next, the four cases will be described and assessed in detail.

**Table 1.** Four cases as regards the target of prediction.

| Case | Target of prediction |
|---|---|
| 1 | Long-term mean in the TPE |
| 2 | A new year at the mean of the TPE |
| 3 | Long-term mean at new location (e.g., a farm) |
| 4 | A new year at a new location (e.g., a farm) |

**Case 1**: Assume that the objective is to obtain an estimate of the *long-term mean in the TPE*. In this case, we evaluate the fitted regression at the unconditional expectation of the EC in the TPE, given by

$$\xi = E(x) = \mu_x, \tag{32}$$

i.e. the regression term is $\gamma_i \mu_x$. As this is the prediction at a long-term mean, there is no uncertainty associated with the value we use for the EC, except that the mean $\mu_x$ needs to be estimated, hence $\phi_{x(i)} = 0$.

**Case 2**: Next, consider the case where prediction *for a new year at the mean of the TPE* is needed. In this case, we would want to replace $x$ by the mean in the TPE in the new year $m_0$, given by

$$\xi_{m_0} = E(x|m_0) = \mu_x + Y_{x(m_0)}. \tag{33}$$

If the covariate is only observed next year, however, we can merely use the long-term mean $\xi$ of $x$ in the TPE, i.e. $\xi = \mu_x$, which deviates from $\xi_{m_0}$ by the amount $Y_{x(m_0)}$. This deviation is also unknown at the time the prediction is needed, but if we have historical data on the EC, we can estimate the variance $var\left(Y_{x(m_0)}\right) = \sigma^2_{x(Y)}$, and hence $\phi_{x(i)} = \gamma_i^2 \sigma^2_{x(Y)}$.

**Case 3:** If we want to estimate the *long-term mean at new location* $l_0$, e.g. a farm, we use the long-term mean of the EC at the new location:

$$\xi_{l_0} = E(x|l_0) = \mu_x + L_{x(l_0)}. \quad (34)$$

Note that in this case, the model in (31) will have a fixed location effect $L_{x(l)}$ because historical data is available for the target location. As we are estimating a long-term mean, we have $\phi_{x(i)} = 0$. Note that estimating $\xi_{l_0}$ requires long-term EC data to be available for the new location.

**Case 4:** Finally, assume that the objective is prediction of a *mean for a new location in a new year*. As in the previous case, model in (31) will have fixed location effect $L_{x(l)}$ and we use the long-term mean $\xi_{l_0}$of the EC at the new location because the values of the EC at the new location for the future year will be unavailable when the prediction is needed. However, this time $\phi_{x(i)} = \gamma_i^2\left(\sigma_{x(Y)}^2 + \sigma_{x(LY)}^2\right)$.

In many studies assessing the predictive benefit of using EC, a leave-one-environment-out cross-validation (CV) strategy is employed by which the left-out environment is used to validate the predictions obtained from a model fitted to the remaining environments. In this kind of CV, the observed EC values for the left-out environment are plugged into the fitted model to obtain a prediction. This means that these predictions for the left out $lm$-th environment use $\xi = x_{lm}$. It should be pointed out than this prediction scenario does not correspond to any of the four cases defined above. Using CV with $\xi = x_{lm}$ certainly can inform about the predictive potential of EC, but we contend that it does not represent a realistic prediction scenario occurring in practice. A useful modification of leave-one-environment-out CV with multi-year and multi-location data mimic the Case 4 by using $\xi = \xi_{l_0}$in eq. (32) for predictions into the left-out environment, involving the $l_0$-th location and a new year $m_0$.

### 3.2 Estimating the prediction variance

The overall prediction variance $\upsilon_i$ associated with the regression term $\gamma_i x$ has two components. The first arises from the fact that both $\gamma_i$ and $\xi$ in the product $\gamma_i\xi$ need to be estimated. The second corresponds to $\phi_{x(i)} = \gamma_i^2\sigma_x^2$, i.e. to the fact that $x$ is replaced by $\xi$, from which it may deviate. First, we consider estimation of the contribution $\gamma_i\xi$ to the predicted mean. We generally assume that EC and yield data, conditional on the EC, are independent. The estimator, being a product of two independent random variables $\hat{\gamma}_i$ and $\hat{\xi}$ has variance (Brown & Alexander 1991; Goodman 1960)

$$var\left(\hat{\gamma}_i\hat{\xi}\right) = \gamma_i^2\, var\left(\hat{\xi}\right) + \xi^2\, var(\hat{\gamma}_i) + var(\hat{\gamma}_i)\, var\left(\hat{\xi}\right). \quad (35)$$

Note that the naïve plug-in estimator of $\gamma_i^2$, $\hat{\gamma}_i^2$, is biased. To see this, assume that the estimator $\hat{\gamma}_i$ is unbiased, i.e. $E(\hat{\gamma}_i) = \gamma_i$. Then from the definition of the variance (Rice 1995, p.124)

$$var(\hat{\gamma}_i) = E(\hat{\gamma}_i^2) - [E(\hat{\gamma}_i)]^2 = E(\hat{\gamma}_i^2) - \gamma_i^2. \quad (36)$$

It emerges from eq. (36) that $E(\hat{\gamma}_i^2) = \gamma_i^2 + var(\hat{\gamma}_i)$, establishing the bias. Hence, we may estimate $\gamma_i^2$ by $\tilde{\gamma}_i^2 = \hat{\gamma}_i^2 - var(\hat{\gamma}_i^2)$ and $\xi^2$ by $\tilde{\xi}^2 = \hat{\xi}^2 - var\left(\hat{\xi}\right)$, where for simplicity we make no distinction in notation between $var\left(\hat{\xi}\right)$ and $var(\hat{\gamma}_i)$ and their estimators that need to be used in practice. Hence, the estimator of (35) is

$$est.var\left(\hat{\gamma}_i\hat{\xi}\right) = \hat{\gamma}_i^2\, var\left(\hat{\xi}\right) + \hat{\xi}^2\, var(\hat{\gamma}_i) - var(\hat{\gamma}_i)\, var\left(\hat{\xi}\right). \quad (37)$$

Next, consider the estimation of $\phi_{x(i)} = \gamma_i^2\sigma_x^2$, where $\sigma_x^2$ represents the variance of a random effect in model (31), or a sum of more than one such variance component, depending on the case considered in Section 3.1 (Cases 1 to 4). The naïve plug-in estimator $\hat{\gamma}_i^2\hat{\sigma}_x^2$ is biased, and bias may be reduced by using the estimator

$$\tilde{\phi}_{x(i)} = [\hat{\gamma}_i^2 - var(\hat{\gamma}_i)]\hat{\sigma}_x^2. \quad (38)$$

The overall prediction variance $\upsilon_i$ associated with $\gamma_i x$ may be estimated by adding (37) and (38).

### 3.3 Extension to multiple EC and inclusion of the intercept

The linear predictor in eq. (1) can be written as

$$\eta_i = \boldsymbol{\gamma}_i'^T\boldsymbol{x}' \quad , \quad (39)$$

where $\boldsymbol{x}' = \left(1, x_1, \dots, x_p\right)^T$ and $\boldsymbol{\gamma}'_i = \left(\alpha_i, \gamma_{i1}, \dots, \gamma_{ip}\right)^T$. When it comes to prediction, $\boldsymbol{x}'$ will be replaced by its conditional expectation, $\boldsymbol{\xi}' = \left(1, \xi_1, \dots, \xi_p\right)^T$, depending on the particular prediction scenario (Cases 1 to 4; see Section 3.1), leading to the predictor

$$\eta_i = \boldsymbol{\gamma}_i'^T \boldsymbol{\xi}' \quad . \tag{40}$$

Prediction using SC can be investigated analogously by simply replacing $\boldsymbol{x}'$ with $\boldsymbol{z}' = \left(1, z_1, \dots, z_q\right)^T$ but for the sake of brevity this will not be considered here explicitly. The conditional variance of $\boldsymbol{x}'$ for given $\boldsymbol{\xi}'$ will be denoted as $\boldsymbol{\Sigma}_{\boldsymbol{x}'}$, and the associated prediction variance will be

$$\phi_{\boldsymbol{x}'(i)} = \boldsymbol{\gamma}_i'^T \boldsymbol{\Sigma}_{\boldsymbol{x}'} \boldsymbol{\gamma}_i' \quad . \tag{41}$$

To assess the prediction variance, we use a multivariate extension of the two-way model (31) for the covariate vector $\boldsymbol{x}'$:

$$\boldsymbol{x}'_{lm} = \boldsymbol{\mu}_{\boldsymbol{x}'} + \boldsymbol{L}_{\boldsymbol{x}'(l)} + \boldsymbol{Y}_{\boldsymbol{x}'(m)} + (\boldsymbol{LY})_{\boldsymbol{x}'(lm)} \quad , \tag{42}$$

where $\boldsymbol{\mu}_{\boldsymbol{x}'} = \left(1, \mu_{x_1}, \dots, \mu_{x_p}\right)^T$ is the overall mean $\boldsymbol{x}'_{lm}$ and the random-effect vectors are similarly defined with variances $var\left(\boldsymbol{L}_{\boldsymbol{x}'(l)}\right) = \boldsymbol{\Sigma}_{\boldsymbol{x}'(L)}$, $var\left(\boldsymbol{Y}_{\boldsymbol{x}'(m)}\right) = \boldsymbol{\Sigma}_{\boldsymbol{x}'(Y)}$, and $var\left((\boldsymbol{LY})_{\boldsymbol{x}'(lm)}\right) = \boldsymbol{\Sigma}_{\boldsymbol{x}'(LY)}$. Note that in cases 3 and 4 this model will feature fixed location effect $\boldsymbol{L}_{\boldsymbol{x}'(l)}$ instead of random, because we focus on a specific location. We note that the first row and column of these three variance-covariance matrices, which correspond to the intercept, have all entries equal to zero. Moreover, covariates that do not change value over years have corresponding zero entries in $\boldsymbol{\Sigma}_{\boldsymbol{x}'(Y)}$ and $\boldsymbol{\Sigma}_{\boldsymbol{x}'(LY)}$. Hence, all three variance-covariance matrices are singular. With these definitions, we can now consider the four cases. The explicit expressions for $\boldsymbol{\xi}'$, $\boldsymbol{\Sigma}_{\boldsymbol{x}'}$ and $\phi_{\boldsymbol{x}'(i)}$ are given in Table 2.

**Table 2.** Explicit expressions for $\boldsymbol{\xi}'$, $\boldsymbol{\Sigma}_{\boldsymbol{x}'}$, $\phi_{\boldsymbol{x}'(i)}$ and $\upsilon_R$ for the four scenarios (Case 1 to 4) when predicting individual genotypes. Note: contribution of deviations from regression (denoted $\upsilon_R$) and variances that compose it and pertain to years ($Y$), locations ($L$), genotypes ($\alpha$), or interactions ($\alpha Y$, $\alpha L$, $\alpha LY$) are discussed in more detail in Section 3.5 "Contribution of the deviations from regression".

| Case | $\boldsymbol{\xi}'$ | $\boldsymbol{\Sigma}_{\boldsymbol{x}'}$ | $\phi_{\boldsymbol{x}'(i)}$ | $\upsilon_R$ |
|---|---|---|---|---|
| 1 | $\boldsymbol{\mu}_{\boldsymbol{x}'}$ | 0 | 0 | 0 |
| 2 | $\boldsymbol{\mu}_{\boldsymbol{x}'}$ | $\boldsymbol{\Sigma}_{\boldsymbol{x}'(Y)}$ | $\boldsymbol{\gamma}_i'^T \boldsymbol{\Sigma}_{\boldsymbol{x}'(Y)} \boldsymbol{\gamma}_i'$ | $\sigma_Y^2 + \sigma_{\alpha Y}^2$ |
| 3 | $\boldsymbol{\mu}_{\boldsymbol{x}'} + \boldsymbol{L}_{\boldsymbol{x}'(l_0)}$ | 0 | 0 | $\sigma_L^2 + \sigma_{\alpha L}^2$ |
| 4 | $\boldsymbol{\mu}_{\boldsymbol{x}'} + \boldsymbol{L}_{\boldsymbol{x}'(l_0)}$ | $\boldsymbol{\Sigma}_{\boldsymbol{x}'(Y)} + \boldsymbol{\Sigma}_{\boldsymbol{x}'(LY)}$ | $\boldsymbol{\gamma}_i'^T \left(\boldsymbol{\Sigma}_{\boldsymbol{x}'(Y)} + \boldsymbol{\Sigma}_{\boldsymbol{x}'(LY)}\right) \boldsymbol{\gamma}_i'$ | $\sigma_L^2 + \sigma_{\alpha L}^2 + \sigma_Y^2 + \sigma_{\alpha Y}^2 + \sigma_{LY}^2 + \sigma_{\alpha LY}^2$ |

### 3.4 Estimating the overall prediction variance

Again, the overall prediction variance has two components, one arising from the fact that $\eta_i = \boldsymbol{\gamma}_i'^T \boldsymbol{\xi}'$ needs to be estimated and the second corresponding to $\boldsymbol{\Sigma}_{\boldsymbol{x}'}$. First, we consider estimation of the predicted mean $\eta_i = \boldsymbol{\gamma}_i'^T \boldsymbol{\xi}'$. We generally assume that EC and yield data are independent. The estimator, being a product of two random vectors $\hat{\boldsymbol{\gamma}}_i'$ and $\hat{\boldsymbol{\xi}}'$ has total variance

$$var\left(\hat{\boldsymbol{\gamma}}_i'^T \hat{\boldsymbol{\xi}}'\right) = \boldsymbol{\gamma}_i'^T \, var\left(\hat{\boldsymbol{\xi}}'\right) \boldsymbol{\gamma}_i' + \boldsymbol{\xi}'^T \, var\left(\hat{\boldsymbol{\gamma}}_i'\right) \boldsymbol{\xi}' + trace\left[var\left(\hat{\boldsymbol{\gamma}}_i'\right) var\left(\hat{\boldsymbol{\xi}}'\right)\right]. \tag{43}$$

This variance may be regarded as straightforward extension of the result for a product of two scalar random variables (Brown & Alexander 1991; Goodman 1960) (see Appendix B for a derivation). We may estimate $\boldsymbol{\gamma}_i'^T \, var\left(\hat{\boldsymbol{\xi}}'\right) \boldsymbol{\gamma}_i'$ by $\hat{\boldsymbol{\gamma}}_i'^T \, var\left(\hat{\boldsymbol{\xi}}'\right) \hat{\boldsymbol{\gamma}}_i' - trace\left[var\left(\hat{\boldsymbol{\gamma}}_i'\right) var\left(\hat{\boldsymbol{\xi}}'\right)\right]$ and $\boldsymbol{\xi}'^T \, var\left(\hat{\boldsymbol{\gamma}}_i\right) \boldsymbol{\xi}'$ by $\hat{\boldsymbol{\xi}}'^T \, var\left(\hat{\boldsymbol{\gamma}}_i\right) \hat{\boldsymbol{\xi}}' - trace\left[var\left(\hat{\boldsymbol{\gamma}}_i'\right) var\left(\hat{\boldsymbol{\xi}}'\right)\right]$, where for simplicity of notation we make no distinction between $var\left(\hat{\boldsymbol{\xi}}'\right)$ and $var\left(\hat{\boldsymbol{\gamma}}_i'\right)$ and their estimators. Hence, the estimator of (43) is

$$est.var\left(\hat{\boldsymbol{\gamma}}_i'^T \hat{\boldsymbol{\xi}}'\right) = \hat{\boldsymbol{\gamma}}_i'^T \, var\left(\hat{\boldsymbol{\xi}}'\right) \hat{\boldsymbol{\gamma}}_i' + \boldsymbol{\xi}'^T \, var\left(\hat{\boldsymbol{\gamma}}_i'\right) \boldsymbol{\xi}' - trace\left[var\left(\hat{\boldsymbol{\gamma}}_i'\right) var\left(\hat{\boldsymbol{\xi}}'\right)\right]. \tag{44}$$

The estimates of $var\left(\hat{\boldsymbol{\xi}}'\right)$ and $var\left(\hat{\boldsymbol{\gamma}}'_i\right)$ needed in (43) can be obtained from the inverse of the coefficient matrix of the mixed model equations after completion of the residual maximum likelihood estimation of the variance components involved (Searle et al. 1992, p. 276).

Next, consider the estimation of $\boldsymbol{\gamma}_i'^T \boldsymbol{\Sigma}_{x'} \boldsymbol{\gamma}_i'$ in (41). The naïve plug-in estimator $\phi_{x'(i)} = \widehat{\boldsymbol{\gamma}}_i'^T \widehat{\boldsymbol{\Sigma}}_{x'} \widehat{\boldsymbol{\gamma}}_i'$ is biased, and bias may be reduced by using the estimator

$$\tilde{\phi}_{x'(i)} = \widehat{\boldsymbol{\gamma}}_i'^T \widehat{\boldsymbol{\Sigma}}_{x'} \widehat{\boldsymbol{\gamma}}_i' \; - \; trace\left[var(\widehat{\boldsymbol{\gamma}}_i')\, \widehat{\boldsymbol{\Sigma}}_{x'}\right]. \tag{45}$$

### 3.5 Contribution of the deviations from regression

In all four cases, the residual terms $u_j$ and $e_{ij}$ from the model in (2) are partitioned by year and location, i.e. we will use

$$u_{lm} = L_l + Y_m + (LY)_{lm} \text{ and} \tag{46}$$
$$e_{ilm} = (\alpha L)_{il} + (\alpha Y)_{im} + (\alpha LY)_{ilm} \,, \tag{47}$$

where $L$, $Y$ and $\alpha$ denote the factors location, year and genotype, and subscripts $l$ and $m$ index locations and years. All effects in (46) and (47) are independently distributed with constant variance, i.e. $L_l \sim N\,(0, \sigma_L^2)$, $Y_m \sim N\,(0, \sigma_Y^2)$, $(LY)_{lm} \sim N\,(0, \sigma_{LY}^2)$, $(\alpha L)_{il} \sim N\,(0, \sigma_{\alpha L}^2)$, $(\alpha Y)_{im} \sim N\,(0, \sigma_{\alpha Y}^2)$, and $(\alpha LY)_{ilm} \sim N\,(0, \sigma_{\alpha LY}^2)$. In each case, a subset of these variances will contribute to the overall uncertainty of the prediction. The variance of this contribution will be denoted as $\upsilon_R$. The total prediction variance is

$$\upsilon_i = var\left(\widehat{\boldsymbol{\gamma}}_i'^T \widehat{\boldsymbol{\xi}}'\right) + \tilde{\phi}_{x'(i)} + \upsilon_R. \tag{48}$$

The explicit expressions of $\upsilon_R$ for the four cases are given in Table 2. Note that in Case 3 $\upsilon_R = \sigma_L^2 + \sigma_{\alpha L}^2$, because the random effects $L_{l_0}$ and $(\alpha L)_{il_0}$ are unknown for a new location. Also, in Case 4 we have $\upsilon_R = \sigma_L^2 + \sigma_{\alpha L}^2 + \sigma_Y^2 + \sigma_{\alpha Y}^2 + \sigma_{LY}^2 + \sigma_{\alpha LY}^2$, because all location- and year-related effects are unknown for a new location and year. If prediction for several new locations in a new year is required, then the $\upsilon_i$ is computed for every new location separately.

### 3.6 Pairwise differences

To compute $\upsilon$ for the pairwise difference of two genotypes $i$ and $i'$, we replace $\boldsymbol{\gamma}_i'$ with $\boldsymbol{\delta}_{ii'}' = \boldsymbol{\gamma}_i' - \boldsymbol{\gamma}_{i'}'$. In the residual prediction variance for a difference $\upsilon_R$, we may drop $\sigma_L^2$, $\sigma_Y^2$ and $\sigma_{LY}^2$, because the corresponding random effects drop out in the pairwise difference. Applying equations (43), (44), and (47) to pairs replacing $\boldsymbol{\gamma}_i'$ with $\boldsymbol{\delta}_{ii'}' = \boldsymbol{\gamma}_i' - \boldsymbol{\gamma}_{i'}'$, we find after some algebra (see Appendix C) the average total prediction variance of a difference

$$\hat{\bar{\upsilon}}_\delta = trace\left\{\left[\mathbf{P} \otimes \left(var\left(\widehat{\boldsymbol{\xi}}'\right) + \widehat{\boldsymbol{\Sigma}}_{x'}\right)\right] \widehat{\boldsymbol{\gamma}}' \widehat{\boldsymbol{\gamma}}'^T\right\} + trace\left[var(\widehat{\boldsymbol{\gamma}}')\left(\mathbf{P} \otimes \widehat{\boldsymbol{\xi}}' \widehat{\boldsymbol{\xi}}'^T\right)\right] - trace\left\{var(\widehat{\boldsymbol{\gamma}}')\left[\mathbf{P} \otimes \left(var\left(\widehat{\boldsymbol{\xi}}'\right) + \widehat{\boldsymbol{\Sigma}}_{x'}\right)\right]\right\} + \hat{\bar{\upsilon}}_R \;, \tag{49}$$

where $\boldsymbol{\gamma}' = \left(\boldsymbol{\gamma}'^T_1, \ldots, \boldsymbol{\gamma}'^T_I\right)^T$, $\mathbf{P} = 2(I-1)^{-1}[\mathbf{I}_I - \mathbf{K}_I]$ with $\mathbf{K}_I = I^{-1} \mathbf{1}_I \mathbf{1}_I^T$, and the average residual prediction variance of a difference $\bar{\upsilon}_R$ is defined in Table 3.

**Table 3.** Explicit expressions for $\boldsymbol{\xi}'$, $\boldsymbol{\Sigma}_{x'}$ and $\bar{\upsilon}_R$ for the four scenarios (Case 1 to 4) when predicting genotype differences and averaging the uncertainty across pairs

| Case | $\boldsymbol{\xi}'$ | $\boldsymbol{\Sigma}_{x'}$ | $\bar{\upsilon}_R$ |
|---|---|---|---|
| 1 | $\boldsymbol{\mu}_{x'}$ | 0 | 0 |
| 2 | $\boldsymbol{\mu}_{x'}$ | $\boldsymbol{\Sigma}_{x'(Y)}$ | $2\sigma_{\alpha Y}^2$ |
| 3 | $\boldsymbol{\mu}_{x'} + \boldsymbol{L}_{x'(l_0)}$ | 0 | $2\sigma_{\alpha L}^2$ |
| 4 | $\boldsymbol{\mu}_{x'} + \boldsymbol{L}_{x'(l_0)}$ | $\boldsymbol{\Sigma}_{x'(Y)} + \boldsymbol{\Sigma}_{x'(LY)}$ | $2(\sigma_{\alpha L}^2 + \sigma_{\alpha Y}^2 + \sigma_{\alpha LY}^2)$ |

## 4 Materials and methods

### 4 .1 Datasets

Long-term multi-environment rice stability trials, described by Rahman et al. (2023) and provided by the Bangladesh Rice Research Institute (BRRI), are used to demonstrate models and methods. These trials include registered varieties from two distinct breeding programs: irrigated winter (dry season) rice and rainfed summer (monsoon season) rice. The two datasets are referred to as *winter rice* and *summer rice*, respectively, and were analyzed separately. They contain yield observations (t/ha) obtained from randomized complete block design with three blocks, replicated across 8 (summer rice) and 9 (winter rice) locations per year, between 2001 and 2022. Over time, the number of summer rice varieties increased from 16 to 45 and the number of winter rice varieties increased from 18 to 42, as new varieties were added while retaining older ones. Varieties tested in fewer than two years were excluded.

### 4.2 Covariates

Weather data were sourced from the AgERA5 database (Boogaard et al. 2020) using the cdsapi application programming interface in Python 3.11.7 (Van Rossum & Drake 2009), and decoded using the ag5Tools library (Brown et al. 2023) in R (R Core Team 2021). Weather covariates were aggregated by growing season. A total of 8 covariates was used for each dataset. A full description of selected covariates is provided in Appendix D.

### 4.3 Models

Both rice datasets were analyzed using a two-stage approach. In the first stage, the genotype means and associated variances were estimated by fitting a model (Rahman et al. 2023):

$$y_{il} = \mu + b_l + g_i + e_{il} , \quad (50)$$

Where $y_{il}$ is the observation of the $i$-th genotype in the $l$-th block, $\mu$ is the fixed intercept, $b_l$ is the $l$-th fixed block effect, $g_i$ is the $i$-th fixed genotype effect, and $e_{il}$ is the independent and identically distributed error associated with the $i$-th genotype and $l$-th block. Inverse variances of adjusted genotype means were used in the second stage of the analysis as diagonal weights. Comparison of models described in Section 2 is done based on their performance in the second stage. All models have fixed overall intercept and fixed covariate slopes, and all the other effects are treated as random. The main genotypic effect $\alpha$ and terms $u_j$ and $e_{ij}$, partitioned by year and location as in (46) and (47), appear in every model unchanged. Therefore, the models differ solely in how covariates are incorporated:

- Baseline model without genotype-specific covariate response.
- Environmental kernel model with the kernel matrix derived from observed covariates on a per-environment basis (15).
- RRR with the reduced rank variance-covariance structure of rank one (RRR1) or two (RRR2) assigned to genotype-specific covariate slopes and the genotypic main effect (18).
- RFR with the unstructured variance-covariance structure assigned to genotype-specific covariate slopes and the genotypic main effect.
- Unstructured regression with synthetic covariates (FW-US) fitted similar to RFR, except that synthetic covariates were used instead of observed. Synthetic covariates were derived as in (24) using the method similar to one described for the Extended Finlay-Wilkinson regression in Piepho & Blancon (2023, Section 5.2, Eq. 18), except that $\alpha L$, $\alpha Y$, and $\alpha LY$ effects were included and fitted as random, as it improved the quality of synthetic covariates. Only one (FW1-US) and two (FW2-US) synthetic covariates were obtained, as it was shown sufficient in related literature (Piepho & Blancon 2023; Tadese et al. 2024).

A side-by-side illustration of these models in simplified notation is available in Appendix E.

In the case of the Kernel model, the number of covariates can be much larger than the number of the environments, and thus the fixed covariate slope cannot be fitted. Because this possibility exists, we investigated it for every model and termed such models "without the main EC effect" in tables and the text.

All models were fitted in ASReml-R 4.2 (Butler et al. 2023) in R programming language version 4.5.1 enhanced by the oneAPI Intel Math Kernel Library version 2024.0.

### 4.4 Model performance evaluation

The models were evaluated through a comparison of the model fit with the entirety of the data, and through two CV scenarios.

The model fit was evaluated based on several criteria, including the number of parameters (accounting for those associated with synthetic covariates, if applicable), the log-likelihood (LogLik) of the fitted model, the Akaike Information Criterion (AIC), the variance components and their percentage change relative to the baseline model. The AIC values were calculated using a method proposed by Verbyla (2019) implemented in the infoCriteria function from the asremlPlus package (Brien 2024), and then adjusted for the number of parameters involved in synthetic covariates:

$$AIC = -2 * LogLik + 2 * (VarP + FixedP + SP) \quad (51)$$

where $LogLik$ is the full likelihood, $VarP$ is the number of estimated variance parameters, $FixedP$ is the number of estimated fixed parameters, and $SP$ is the number of parameters involved in synthetic covariates. The $SP$ of FW1-US was 8, while for FW2-US, the $SP$ was 15. The $SP$ of other models was 0. Parameters fixed by ASReml-R during model fitting were nevertheless included in the total number of parameters. Parameters that were fixed by design were not included in the AIC (i.e., the residual variance parameter fixed at 1 to correctly provide weights from the first stage analysis, and the first loading of the second-order reduced rank variance-covariance structure fixed at 0).

The CV was used to evaluate the performance of models using various criteria: Pearson's correlation coefficient (PCC) between predicted and observed values, mean squared prediction error (MSPE), mean squared error of predicted differences (MSEPD), mean variance of predictions (MVP), and mean variance of predicted differences (VPD).

MSPE is defined as:

$$MSPE_j = \frac{\sum_{i=1}^{I}(y_{ij} - \hat{y}_{ij})^2}{I}, \quad (52)$$

where $y_{ij}$ is the observed phenotypic value of the $i$-th genotype in the $j$-th environment, $\hat{y}_{ij}$ is the respective predicted phenotypic value, and $I$ is the total number of genotypes. Environment-specific means are then aggregated as means or medians.

MSEPD is defined as (Piepho 1998):

$$MSEPD_j = \frac{2\sum_{i=1}^{I}(f_{ij} - \bar{f}_{\bullet j})^2}{(I-1)}, \quad (53)$$

where $f_{ij} = y_{ij} - \hat{y}_{ij}$ similar to MSPE, and the $\bar{f}_{\bullet j}$ is the mean of $f_{ij}$ in the $j$-th environment. Environment-specific means are then aggregated as means or medians.

Variance of prediction and VPD were calculated as in (48) and (49) respectively. The variance of prediction was averaged across genotypes within an environment to get the MVP. Such environment-specific values of MVP and VPD were aggregated as means or medians.

The first CV scenario is a common leave-one-environment-out (LOEO) approach, where known covariate values are used to predict the phenotypic response in the left-out environment. The second scenario mimics prediction in an unseen environment by leaving an entire year and location out (LYLO). In this case, prediction is performed for the left-out location in the left-out year using the mean covariate values from available years for that location. In LYLO CV, the effects $Y$, $L$, $YL$, $\alpha L$, $\alpha Y$, and $\alpha LY$ are not available for prediction. In contrast, under the LOEO scenario, only $YL$ and $\alpha LY$ are not available. Covariate value in LYLO CV can be imputed as a simple mean over years in the training set or calculated using a multivariate mixed model as in (42), which is numerically identical in this setting. This model is required to estimate the variances and covariances of EC for the MVP and VPD.

**4.5 Critical points in model fitting**

Covariates were centered and scaled to unit variance only once in LOEO CV, because the table of covariates is the same for every CV split there. In LYLO CV we repeated centering and scaling in every split.

To compute prediction variance using the method suggested in this paper, we need to extract the so-called C-inverse matrix, which contains the variances and covariances of genotypic intercepts and genotype-specific covariate slopes. This matrix is only partially available in ASReml 4.2. Therefore, the approach described by Henderson (1984, Chapter 5) was implemented for both singular and non-singular G matrices. A script illustrating the comparison of the two matrices is available on GitHub.

RRR1, RRR2, and RFR models had sometimes issues with convergence, therefore we precomputed each model with all the data and used the variance component estimates as starting values when there were issues. The model was forced to run additional iterations in case ASReml-R reported convergence, but the LogLik continued to visibly increase. The model fitting was stopped if the LogLik suddenly increased to an unrealistic value or started to decrease.

Running FW1-US and FW2-US models involved two steps: one to extract synthetic covariates, and one to fit them. In both CV scenarios, extraction of synthetic covariates was done in every CV split.

## 5 Results

The results from the winter and summer rice datasets are largely consistent, although winter rice was irrigated, which likely disrupts the connection between plant performance and external weather conditions. For brevity, results related to the winter rice dataset are provided in the Online Resource (Supplementary Information). Nevertheless, this section highlights the differences between the results from the two datasets.

The fit of models on complete data is summarized in Table 4. The models with the main EC effect always had more parameters and lower LogLik than their counterparts, which did not translate into a notable AIC reduction. The RFR had the highest number of parameters, the smallest LogLik, but AIC on par with RRR1 model that approximates RFR with less parameters. The smallest AIC was achieved by RRR2 model, followed by FW1-US and FW2-US.

**Table 4.** Model fit (summer rice) featuring the total number of parameters in the model (including those involved in synthetic covariates), full log-likelihood, and Akaike Information Criterion (AIC). Displayed values were rounded. Baseline – model without genotype-covariate interactions, Kernel – model with an environmental kernel matrix, RRR1 and RRR2 – reduced rank regression of rank one and two with observed covariates, RFR - random factorial regression with observed covariates, FW1-US and FW2-US – random factorial regression with one and two synthetic covariates respectively.

| **With the main EC effect** | | | | **Without the main EC effect** | | | |
|---|---|---|---|---|---|---|---|
| **Model** | **Parameters** | **LogLik** | **AIC** | **Model** | **Parameters** | **LogLik** | **AIC** |
| Baseline | 16 | -459 | 949 | Baseline | 8 | -465 | 947 |
| Kernel | 17 | -449 | 931 | Kernel | 9 | -455 | 929 |
| RRR1 | 24 | -443 | 933 | RRR1 | 16 | -450 | 932 |
| RRR2 | 32 | -412 | 888 | RRR2 | 24 | -420 | 887 |
| RFR | 60 | -407 | 935 | RFR | 52 | -415 | 934 |
| FW1-US | 19 | -434 | 905 | FW1-US | 18 | -435 | 907 |
| FW2-US | 30 | -424 | 907 | FW2-US | 28 | -428 | 911 |

The variance components of different models are summarized as percent change relative to the baseline model in Table 5. All variance component values for summer rice are available in Supplementary Information (Table S2). The reduction in the $\alpha L$, $\alpha Y$, and $\alpha LY$ components is linked to the incorporation of environmental covariates. However, the substantial reduction observed in the $\alpha L$ and $\alpha Y$ components is relatively minor compared to the magnitude of the $LY$ and $\alpha LY$ components.

**Table 5.** Variance components percent change relative to the baseline (summer rice). Baseline – model without genotype-covariate interactions, Kernel – model with an environmental kernel matrix, RRR1 and RRR2 – reduced rank regression of rank one and two with observed covariates, RFR – random factorial regression with observed covariates, FW1-US and FW2-US – random factorial regression with one and two synthetic covariates, $L$ – location, $Y$ – year, $\alpha$ – genotype.

| **Models with the main EC effect** | | | | | | | |
|---|---|---|---|---|---|---|---|
| **Component** | **Baseline[a]** | **Kernel** | **RRR1** | **RRR2** | **RFR** | **FW1-US** | **FW2-US** |
| $L$ | 0.0568 | -2.5 | -2.0 | -5.6 | -5.5 | 78.8 | -5.8 |
| $Y$ | 0.0262 | 3.3 | -1.9 | 9.6 | 9.6 | -9.6 | 41.5 |
| $\alpha$ | 0.2422 | 0.7 | -4.7 | -4.2 | -5.4 | -3.4 | -3.2 |
| $LY$ | 0.4434 | 0.2 | 0.3 | 0.7 | 0.6 | 1.0 | -0.1 |
| $\alpha L$ | 0.0307 | -26.7 | -12.6 | -27.6 | -39.3 | -19.1 | -29.0 |
| $\alpha Y$ | 0.0142 | -24.4 | -16.0 | -50.4 | -58.0 | -31.7 | -32.8 |
| $\alpha LY$ | 0.2695 | -0.1 | 0.4 | -0.6 | -1.0 | -1.0 | -1.1 |
| **Models without the main EC effect** | | | | | | | |
| **Component** | **Baseline[a]** | **Kernel** | **RRR1** | **RRR2** | **RFR** | **FW1-US** | **FW2-US** |
| $L$ | 0.1269 | 0.5 | 2.1 | 2.6 | 3.0 | 0.9 | 2.4 |
| $Y$ | 0.0250 | 5.3 | 9.8 | 28.4 | 28.0 | 20.2 | 18.9 |

| $\alpha$ | 0.2430 | 0.8 | -4.2 | -3.5 | -4.7 | -3.2 | -2.9 |
|---|---|---|---|---|---|---|---|
| $LY$ | 0.4441 | 0.2 | 0.4 | 0.8 | 0.7 | 0.4 | 0.7 |
| $\alpha L$ | 0.0307 | -26.5 | -12.5 | -28.0 | -39.6 | -19.1 | -29.1 |
| $\alpha Y$ | 0.0142 | -24.2 | -16.0 | -50.1 | -57.8 | -31.7 | -32.9 |
| $\alpha LY$ | 0.2695 | -0.1 | 0.4 | -0.6 | -1.0 | -1.0 | -1.0 |

[a] represents the actual values from the Baseline model, the rest are in percent relative to it.

The results of LOEO and LYLO CVs are reported in Table 6. We also provided the medians of PCC, MSEPD, and MSPE in the Supplementary Information (Table S7), because their distributions are skewed.

The models that use EC generally outperform the baseline in both LOEO and LYLO scenarios. This is seen from the mean performance of models displayed in Table 6. Only RRR1 and FW2-US are lagging behind the baseline in a few cases. Notably, all models benefited from inclusion of the mean regression on EC in LYLO CV, but not in LEO CV.

However, if the CV statistics are aggregated as medians (Supplementary Information, Table S7), then some of the trends differ. This is due to the left-skewed distribution of PCC values, and those right-skewed of MSEPD and MSPE. If medians are considered, then the very bad CV folds are discounted, and thus the model ranking changes given overall low magnitude of differences between them.

The same applies to the winter rice dataset (Supplementary Information), except that most models perform worse than the baseline in LYLO CV by both means and medians.

**Table 6.** Leave-one-environment-out (LOEO) and leave-one-year-and-location-out (LYLO) cross-validation means (summer rice). PCC – Pearson's correlation coefficient, MSEPD – mean squared error of predicted difference, MSPE – mean squared prediction error, Baseline – model without genotype-covariate interactions, Kernel – model with an environmental kernel matrix, RRR1 and RRR2 – reduced rank regression of rank one and two with observed covariates, RFR – random factorial regression with observed covariates, FW1-US and FW2-US – random factorial regression with one and two synthetic covariates.

| Type | Model | Mean PCC | | Mean MSEPD | | Mean MSPE | |
|---|---|---|---|---|---|---|---|
| | | LOEO | LYLO | LOEO | LYLO | LOEO | LYLO |
| With the main EC effect | Baseline | 0.618 | 0.589 | 0.753 | 0.807 | 0.887 | 1.01 |
| | Kernel | 0.620 | 0.595 | 0.750 | 0.795 | 0.887 | 0.95 |
| | RRR1 | 0.617 | 0.589 | 0.754 | 0.805 | 0.889 | 0.961 |
| | RRR2 | 0.621 | 0.591 | 0.751 | 0.805 | 0.890 | 0.967 |
| | RFR | 0.620 | 0.591 | 0.751 | 0.803 | 0.890 | 0.965 |
| | FW1-US | 0.622 | 0.591 | 0.750 | 0.804 | 0.872 | 0.997 |
| | FW2-US | 0.619 | 0.590 | 0.759 | 0.805 | 0.890 | 0.962 |
| Without the main EC effect | Baseline | 0.618 | 0.589 | 0.753 | 0.807 | 0.868 | 1.02 |
| | Kernel | 0.620 | 0.595 | 0.750 | 0.795 | 0.868 | 1.02 |
| | RRR1 | 0.617 | 0.589 | 0.754 | 0.805 | 0.872 | 1.03 |
| | RRR2 | 0.621 | 0.592 | 0.751 | 0.805 | 0.874 | 1.04 |
| | RFR | 0.620 | 0.591 | 0.751 | 0.802 | 0.874 | 1.03 |
| | FW1-US | 0.622 | 0.591 | 0.750 | 0.804 | 0.871 | 1.03 |
| | FW2-US | 0.619 | 0.590 | 0.759 | 0.805 | 0.876 | 1.03 |

The comparison of model-based MVP and VPD with their CV-based counterparts (MSPE and MSEPD) is given in Table 7 and Table 8. The median MVP quite well approximates ranking of models according to the mean MSPE. Mean and median VPD does not differ much, and also approximates CV model performance well except for FW2-US. In the winter rice dataset, despite models not outperforming the baseline, the MVP and VPD estimates favor more complex models.

**Table 7.** Variance of the prediction (MVP) and mean squared prediction error (MSPE) from leave-one-year-and-location-out cross-validation (summer rice). Baseline – model without genotype-covariate interactions, Kernel – model with an environmental kernel matrix, RRR1 and RRR2 – reduced rank regression of rank one and two with observed covariates, RFR – random factorial regression with observed covariates, FW1-US and FW2-US – random factorial regression with one and two synthetic covariates.

| Type | Model | MSPE | | MVP | |
|---|---|---|---|---|---|
| | | Mean | Median | Mean | Median |
| With the main EC effect | Baseline | 1.01 | 0.656 | 0.956 | 0.930 |
| | RRR1 | 0.96 | 0.660 | 0.92 | 0.896 |
| | RRR2 | 0.97 | 0.665 | 0.93 | 0.902 |
| | RFR | 0.96 | 0.665 | 0.92 | 0.896 |
| | FW1-US | 1.00 | 0.642 | 0.92 | 0.919 |
| | FW2-US | 0.96 | 0.625 | 0.90 | 0.898 |
| Without the main EC effect | Baseline | 1.02 | 0.644 | 0.941 | 0.945 |
| | RRR1 | 1.03 | 0.650 | 0.95 | 0.950 |
| | RRR2 | 1.04 | 0.672 | 0.96 | 0.956 |
| | RFR | 1.03 | 0.678 | 0.95 | 0.952 |
| | FW1-US | 1.03 | 0.671 | 0.94 | 0.942 |
| | FW2-US | 1.03 | 0.672 | 0.94 | 0.941 |

**Table 8.** Variance of the predicted difference (VPD) and mean squared error of predicted difference (MSEPD) from the leave-one-year-and-location-out cross-validation (summer rice). Baseline – model without genotype-covariate interactions, Kernel – model with an environmental kernel matrix, RRR1 and RRR2 – reduced rank regression of rank one and two with observed covariates, RFR – random factorial regression with observed covariates, FW1-US and FW2-US – random factorial regression with one and two synthetic covariates. VPD for the baseline model was taken from the standard ASReml-R output.

| Type | Model | MSEPD | | VPD | |
|---|---|---|---|---|---|
| | | Mean | Median | Mean | Median |
| With the main EC effect | Baseline | 0.807 | 0.699 | 0.652 | 0.651 |
| | RRR1 | 0.805 | 0.709 | 0.649 | 0.647 |
| | RRR2 | 0.805 | 0.720 | 0.646 | 0.639 |
| | RFR | 0.803 | 0.720 | 0.632 | 0.632 |
| | FW1-US | 0.804 | 0.717 | 0.635 | 0.629 |
| | FW2-US | 0.805 | 0.716 | 0.627 | 0.624 |
| Without the main EC | Baseline | 0.807 | 0.699 | 0.652 | 0.650 |
| | RRR1 | 0.805 | 0.711 | 0.649 | 0.647 |
| | RRR2 | 0.805 | 0.720 | 0.644 | 0.638 |
| | RFR | 0.802 | 0.720 | 0.630 | 0.631 |
| | FW1-US | 0.804 | 0.716 | 0.634 | 0.629 |
| | FW2-US | 0.805 | 0.716 | 0.627 | 0.624 |

## 6 Discussion

**Behavior of the variance components:** The inclusion of the main EC effect into the model resulted in reduction of the variance component for location in both datasets. The interpretation would be that EC explain a part of the location variation, but not all of it – in that case the variance component would be zero. Notably, FW1-US, that uses only one synthetic covariate in the fixed part, had smaller reduction of the location variance component. One explanation could be that a single synthetic covariate can capture underlying variation in only one dimension, and thus may lag behind models that use several covariates.

**Model performance in different CV scenarios:** There is some performance improvement of most models over the baseline, however the magnitude of this improvement is small. This may seem in contradiction with the large amount of the variance components reduction shown in Table 5, however, it must be borne in mind that reduction of variance components depends only on the structure of covariates in relation to the dataset, and does not necessarily show a model's ability to predict new data points (Sorensen 2023, p. 269). It would be hard to pinpoint the exact reason, but it can be an interplay of lack of fine-scale linkage of covariates to the developmental stages of rice with inability to capture micro-climatic conditions of irrigation in rice using open weather data sources. We hope that with a better data resolution such models would bring a more substantial improvement. The differences between the two CV scenarios also highlight that the validation based on observed EC values primarily implies predictive ability for untested environments from the genotype's perspective, whereas model performance in unseen environments, which would mean any environment in the future, should include removal of EC information to reflect the uncertainty of the unobserved year.

**Fitting RRR and RFR models:** RRR and RFR models can have a large number of effects when many EC are fitted, leading to singularities, unstable convergence, or overfitting. Some of the issues can be resolved by getting variance parameters from a related model (e.g., all design-related components, genotypic variance and residual genetic terms $\alpha L$, $\alpha Y$, and $\alpha LY$) and using these estimates as starting values. Fitting of RFR can be made easier by pre-computing variance-covariance matrix from a simpler RRR1 model, and using these as starting values of the unstructured variance-covariance matrix. For prediction in an unseen environment (e.g. LYLO CV) one can also remove the $\alpha L$ and $\alpha Y$ residual genetic terms and simply fit $\alpha LY$, in case there is not much information to estimate the former.

**Comparison with other studies:** We explored model performance in a setting where exact EC values are unavailable for the next – or generally unseen – year. This scenario is less common than prediction using observed covariates for model validation, as often done with linear models (Hu et al. 2025; Nguyen et al. 2023; Tolhurst 2023) or machine and deep learning models (Yu et al. 2025; Zou et al. 2025). While this provides valuable insight for model comparison and often presents an optimistic view of the potential of EC use, it does not reflect the reality that exact EC values are inherently unknown for future years. Several studies have addressed this issue in different ways – for example, Costa-Neto et al. (2023) used quantile-based long-term, location-level environmental patterns to derive environmental weights, and Gillberg et al. (2019) computed the median of predictions from years with historical weather data to generate predictions for the unseen year. In this context, our work contributes by developing a model-based framework for quantifying prediction uncertainty under unknown future EC values. Regarding predictive performance improvement, although our results are relatively moderate for both LOEO and LYLO cross-validation, many of the cited studies report considerable improvements over their respective baselines, offering a generally optimistic outlook for the future application of EC in MET modeling.

**Differences of distributions of model-based and CV-based values:** The model-based MVP was close to the mean MSPE, and model-based VPD approached median MSEPD. A slight advantage of more complex models in the winter rice dataset by MVP and VPD can be attributed to the greater reduction in variance components achieved through more complex covariance structures. All the CV-based statistics have skewed distributions with quite a few outliers that negatively influence the mean PCC, MSEPD and MSPE (Supplementary Information, Fig. S2-S5). In addition, the distribution of model-based values has much less spread, which makes it complicated to efficiently compare the two visually. It would be desirable to test the means or medians of the related model-based and CV-based values. This, unfortunately, is not possible using known tests, because the assumption of independence of pairs is violated (Schulz-Kümpel et al. 2024). There are several considerations which could help explain the observed differences between those values (Table 7, Table 8). One of them is that we do not account for the uncertainty related to the estimation of variance components. The other one is the fact that the model-based MVP assumes that the fitted model is the correct one. Hence, it misses the bias due to model misspecification (Sorensen 2023), which is a general phenomenon for model-based prediction uncertainty in mixed models.

**Differences between the partial C-inverse matrix of ASReml and our full C-inverse matrix:** We observed minor numerical differences between the C-inverse obtained using method from Henderson (1984, Chapter 5) and the C-inverse obtained from the ASReml-R output. These small discrepancies are likely due to rounding differences in the underlying matrix computations of ASReml-R and our implementation. Additionally, the C-inverse matrix from ASReml-R contained many zeros, whereas our version was fully populated. A script illustrating the comparison between the two matrices is available in the GitHub repository associated with this paper.

**Extension to other sources of data:** The model-based method for uncertainty estimation has applications beyond prediction for an unseen year. It is often overlooked that public EC datasets provide interpolated data, which introduces an inherent estimation error that must be accounted for. Unfortunately, this estimation uncertainty is typically not available in those databases. We therefore hope that in the future, not only EC rasters but also corresponding rasters of EC estimation variance will be made available to enable statisticians to produce more realistic assessments of prediction uncertainty. Another important consideration is that the covariates used in this study are environment-specific. However, it is possible in theory to create genotype specific covariates based on developmental stages. In such case the covariate vector $\boldsymbol{x}$ has to be indexed by genotypes. RFR, RRR, FW-US models would work in such circumstances, with the only change being the notation. However, it would not be possible to create the kernel matrix in the same way as was suggested here, and a different approach should be implemented (e.g., Jarquín et al. 2014). Another minor modelling change would touch the multivariate model to get $\boldsymbol{\Sigma}_{x'}$, because the $(LY)_{lm}$ will not be confounded with the residual error, and the latter has to be modelled additionally. Genomic information can be incorporated into all the models through a genomic relationship matrix. The computation of prediction uncertainty is also generalizable to this case.

## 7 Conclusions

In conclusion, integrating environmental covariates through regression models can improve the prediction of genotype performance in new environments. Among the methods compared, FW1-US and environmental kernel model emerged as promising models, balancing prediction accuracy and model parsimony. We also demonstrated a novel approach to estimate prediction uncertainty, which can help quantify confidence in varietal recommendations when EC must be estimated. Our findings underscore the need in high-quality, high-resolution environmental data to enable more reliable selection decisions. Future research may explore adding non-linearity in these models or incorporate genomic data to further enhance predictions, but the linear mixed models reviewed here provide a strong and interpretable foundation for EC-based prediction.

## Statements and declarations

### Funding

This work was supported by the German Research Foundation (DFG) through grant PI 377/22-2.

### Competing interests

The authors have no relevant financial or non-financial interests to disclose.

### Data availability

The rice dataset of the BRRI can be requested from the authors.

Example R code for the model fitting and the cross-validation can be found on GitHub: https://github.com/m-hrachov/Regression-approaches-for-modelling-genotype-environment-interaction

### Acknowledgement

We would like to thank the BRRI for their dedicated work in collecting and maintaining the long-term rice stability trials data, and for providing these data for the purposes of this study. We also thank the reviewers for their valuable insights and constructive suggestions, which greatly contributed to improving this paper.

**Appendix A**

Number of parameters

$$\boldsymbol{\Sigma}_{RFR} = \begin{pmatrix} \sigma_{\alpha}^2 & \sigma_{\alpha\gamma_1} & \cdots & \sigma_{\alpha\gamma_p} \\ \sigma_{\alpha\gamma_1} & \sigma_{\gamma_1}^2 & \cdots & \sigma_{\gamma_1\gamma_p} \\ \vdots & \vdots & \ddots & \vdots \\ \sigma_{\alpha\gamma_p} & \sigma_{\gamma_1\gamma_p} & \cdots & \sigma_{\gamma_p}^2 \end{pmatrix}$$

$$\boldsymbol{\Sigma}_{RRR(1)} = \boldsymbol{\Lambda}\boldsymbol{\Lambda}^T = \begin{pmatrix} \lambda_{\alpha} \\ \lambda_{\gamma_1} \\ \vdots \\ \lambda_{\gamma_p} \end{pmatrix} \begin{pmatrix} \lambda_{\alpha} & \lambda_{\gamma_1} & \cdots & \lambda_{\gamma_p} \end{pmatrix} = \begin{pmatrix} \lambda_{\alpha}\lambda_{\alpha} & \lambda_{\alpha}\lambda_{\gamma_1} & \cdots & \lambda_{\alpha}\lambda_{\gamma_p} \\ \lambda_{\alpha}\lambda_{\gamma_1} & \lambda_{\gamma_1}\lambda_{\gamma_1} & \cdots & \lambda_{\gamma_1}\lambda_{\gamma_p} \\ \vdots & \vdots & \ddots & \vdots \\ \lambda_{\alpha}\lambda_{\gamma_p} & \lambda_{\gamma_1}\lambda_{\gamma_p} & \cdots & \lambda_{\gamma_p}\lambda_{\gamma_p} \end{pmatrix} \approx \boldsymbol{\Sigma}_{RFR}$$

$$\boldsymbol{\Sigma}_{Kernel} = \begin{pmatrix} \sigma_{\alpha}^2 & 0 \\ 0 & \mathbf{I}_p\sigma_{\gamma}^2 \end{pmatrix} = \begin{pmatrix} \sigma_{\alpha}^2 & 0 & 0 & \cdots & 0 \\ 0 & \sigma_{\gamma}^2 & \cdots & \cdots & 0 \\ \vdots & \vdots & \sigma_{\gamma}^2 & \cdots & 0 \\ \vdots & \vdots & \vdots & \ddots & \vdots \\ 0 & 0 & 0 & \cdots & \sigma_{\gamma}^2 \end{pmatrix}$$

**Figure A1:** Relation of the RFR, RRR (of rank 1 as example), and the Kernel model. In red – parameters that need to be estimated. In blue – parameters that are duplicates or can be derived from estimated ones. $\boldsymbol{\Sigma}_{RFR}$ – unstructured variance-covariance for the genotype main effect and genotype-specific covariate slopes. $\boldsymbol{\Sigma}_{RRR(1)}$ – variance-covariance for the genotype main effect and genotype-specific covariate slopes of the RRR model of rank one as an example. $\boldsymbol{\Sigma}_{Kernel}$ – variance-covariance for the genotype main effect and genotype-specific covariate slopes of the environmental kernel model. $\sigma_{\alpha}^2$, $\sigma_{\gamma}^2$, $\sigma_{\alpha\gamma}$, $\sigma_{\gamma\gamma}$ represent variances for genotypic effect

($\alpha$), EC ($\gamma$), and covariances between genotypic main effect and an EC, or between two EC. The number of EC goes from 1 to $p$. $\lambda$ represents factor loadings for the genotypic main effect and EC.

**Appendix B**

By the law of total variance (Searle et al. 1992, p.461)

$$var\left(\widehat{\boldsymbol{\gamma}}_i'^T \widehat{\boldsymbol{\xi}}'\right) = E_{\widehat{\xi}'}\left[var\left(\widehat{\boldsymbol{\gamma}}_i'^T \widehat{\boldsymbol{\xi}}' | \widehat{\boldsymbol{\xi}}'\right)\right] + var\left[E\left(\widehat{\boldsymbol{\gamma}}_i'^T \widehat{\boldsymbol{\xi}}' | \widehat{\boldsymbol{\xi}}'\right)\right] \widehat{\boldsymbol{\xi}}' \ . \tag{B1}$$

For the conditional variance we find

$$var\left(\widehat{\boldsymbol{\gamma}}_i'^T \widehat{\boldsymbol{\xi}}' | \widehat{\boldsymbol{\xi}}'\right) = \widehat{\boldsymbol{\xi}}'^T \, var(\widehat{\boldsymbol{\gamma}}_i') \widehat{\boldsymbol{\xi}}' \tag{B2}$$

and hence from results on the variance of quadratic forms (Searle et al. 1992, p.466)

$$E_{\widehat{\xi}'}\left[var\left(\widehat{\boldsymbol{\gamma}}_i'^T \widehat{\boldsymbol{\xi}}' | \widehat{\boldsymbol{\xi}}'\right)\right] = \boldsymbol{\xi}'^T \, var(\widehat{\boldsymbol{\gamma}}_i') \boldsymbol{\xi}' + trace\left[var(\widehat{\boldsymbol{\gamma}}_i') \, var\left(\widehat{\boldsymbol{\xi}}'\right)\right] \ . \tag{B3}$$

For the conditional expectation we find

$$E\left(\widehat{\boldsymbol{\gamma}}_i'^T \widehat{\boldsymbol{\xi}}' | \widehat{\boldsymbol{\xi}}'\right) = \widehat{\boldsymbol{\gamma}}_i' \widehat{\boldsymbol{\xi}}' \tag{B4}$$

and hence

$$var_{\widehat{\xi}'}\left[E\left(\widehat{\boldsymbol{\gamma}}_i'^T \widehat{\boldsymbol{\xi}}' | \widehat{\boldsymbol{\xi}}'\right)\right] = \boldsymbol{\gamma}_i'^T var\left(\widehat{\boldsymbol{\xi}}'\right) \boldsymbol{\gamma}_i' \ . \tag{B5}$$

Plugging (B3) and (B5) into (B1), we obtain the total variance in (43).

**Appendix C**

For the pairwise difference of two genotypes $i$ and $i'$, we replace $\boldsymbol{\gamma'}_i$ with $\boldsymbol{\delta'}_{ii'} = \boldsymbol{\gamma}_i' - \boldsymbol{\gamma}_{i'}'$. This This difference can be written

$$\boldsymbol{\delta'}_{ii'} = \boldsymbol{\gamma}_i' - \boldsymbol{\gamma}_{i'}' = \boldsymbol{k}_{ii'}^T \boldsymbol{\gamma}' \ , \tag{C1}$$

where $\boldsymbol{k}_{ii'} = \boldsymbol{c}_{ii'} \otimes \mathbf{I}_{p'}$ with $\boldsymbol{c}_{ii'}$ an $I$-dimensional coefficient vector for the pairwise difference among the two genotypes ($I$ is the total number of genotypes) and $p' = p + 1$ and $\mathbf{I}_{p'}$ the $p'$-dimensional identity matrix. Now evaluating the first term in (44), we may re-arrange

$$\widehat{\boldsymbol{\delta}}_{ii'}^T var\left(\widehat{\boldsymbol{\xi}}'\right) \widehat{\boldsymbol{\delta}}_{ii'} = \widehat{\boldsymbol{\gamma}}'^T \, \boldsymbol{k}_{ii'} var\left(\widehat{\boldsymbol{\xi}}'\right) \boldsymbol{k}_{ii'}^T = trace\left[\boldsymbol{k}_{ii'} var\left(\widehat{\boldsymbol{\xi}}'\right) \boldsymbol{k}_{ii'}^T \widehat{\boldsymbol{\gamma}}' \widehat{\boldsymbol{\gamma}}'^T\right]. \tag{C2}$$

For $i = 1$ and $i' = 2$ this can be written

$$\widehat{\boldsymbol{\delta}}_{12}^T var\left(\widehat{\boldsymbol{\xi}}'\right) \widehat{\boldsymbol{\delta}}_{12} = trace\left[\begin{pmatrix} var\left(\widehat{\boldsymbol{\xi}}'\right) & -var\left(\widehat{\boldsymbol{\xi}}'\right) & \cdots & 0 \\ -var\left(\widehat{\boldsymbol{\xi}}'\right) & var\left(\widehat{\boldsymbol{\xi}}'\right) & \cdots & \vdots \\ \vdots & \vdots & \ddots & \vdots \\ 0 & \cdots & \cdots & 0 \end{pmatrix} \widehat{\boldsymbol{\gamma}}' \widehat{\boldsymbol{\gamma}}'^T \right] \ . \tag{C3}$$

It emerges that the average across pairs takes the form

$$trace\left\{\left[\mathbf{P} \otimes var\left(\widehat{\boldsymbol{\xi}}'\right)\right] \widehat{\boldsymbol{\gamma}}' \widehat{\boldsymbol{\gamma}}'^T\right\}, \tag{C4}$$

where $\mathbf{P} = 2(I-1)^{-1}[\mathbf{I}_I - \mathbf{K}_I]$ with $\mathbf{K}_I = I^{-1} \mathbf{1}_I \mathbf{1}_I^T$. Next, we evaluate the second term in (44), which for the $ii'$-th pair we may re-arrange as

$$\hat{\boldsymbol{\xi}}'^T \; var\big(\hat{\boldsymbol{\delta}}_{ii'}\big)\hat{\boldsymbol{\xi}}' = \hat{\boldsymbol{\xi}}'^T var(\boldsymbol{k}_{ii'}^T\hat{\boldsymbol{\gamma}}')\hat{\boldsymbol{\xi}}' = \hat{\boldsymbol{\xi}}'^T \boldsymbol{k}_{ii'}^T var(\hat{\boldsymbol{\gamma}}')\boldsymbol{k}_{ii'}\hat{\boldsymbol{\xi}}' = \hat{\boldsymbol{\xi}}'^T\big(\boldsymbol{c}_{ii'}^T \otimes \mathbf{I}_{p'}\big)\, var(\hat{\boldsymbol{\gamma}}')\,\big(\boldsymbol{c}_{ii'} \otimes \mathbf{I}_{p'}\big)\hat{\boldsymbol{\xi}}' = trace\big[var(\hat{\boldsymbol{\gamma}}')\,\big(\boldsymbol{c}_{ii'} \otimes \mathbf{I}_{p'}\big)\hat{\boldsymbol{\xi}}'\hat{\boldsymbol{\xi}}'^T\big(\boldsymbol{c}_{ii'}^T \otimes \mathbf{I}_{p'}\big)\big]\,. \tag{C5}$$

For $i = 1$ and $i' = 2$ this can be written

$$\hat{\boldsymbol{\xi}}'^T\, var\big(\hat{\boldsymbol{\delta}}_{12}\big)\hat{\boldsymbol{\xi}}' \;=\; trace\left[var(\hat{\boldsymbol{\gamma}}')\begin{pmatrix} \hat{\boldsymbol{\xi}}'\hat{\boldsymbol{\xi}}'^T & -\hat{\boldsymbol{\xi}}'\hat{\boldsymbol{\xi}}'^T & \cdots & 0 \\ -\hat{\boldsymbol{\xi}}'\hat{\boldsymbol{\xi}}'^T & \hat{\boldsymbol{\xi}}'\hat{\boldsymbol{\xi}}'^T & \cdots & \vdots \\ \vdots & \vdots & \ddots & \vdots \\ 0 & \cdots & \cdots & 0 \end{pmatrix}\right]. \tag{C6}$$

The average across pairs takes the form

$$trace\big[var(\hat{\boldsymbol{\gamma}}')\,\big(\mathbf{P} \otimes \hat{\boldsymbol{\xi}}'\hat{\boldsymbol{\xi}}'^T\big)\big]\,. \tag{C7}$$

For the third term in (44) we find

$$-trace\big[var\big(\hat{\boldsymbol{\delta}}_{ii'}\big)\, var\big(\hat{\boldsymbol{\xi}}'\big)\big] = -trace\big[\boldsymbol{k}_{ii'}^T\, var(\hat{\boldsymbol{\gamma}}')\, \boldsymbol{k}_{ii'}\, var\big(\hat{\boldsymbol{\xi}}'\big)\big] = -trace\big[var(\hat{\boldsymbol{\gamma}}')\, \boldsymbol{k}_{ii'}\, var\big(\hat{\boldsymbol{\xi}}'\big)\, \boldsymbol{k}_{ii'}^T\big]\;, \tag{C8}$$

which for $i = 1$ and $i' = 2$ takes the form

$$-trace\big[var\big(\hat{\boldsymbol{\delta}}_{12}\big)\, var\big(\hat{\boldsymbol{\xi}}'\big)\big] \;=\; -trace\left[var(\hat{\boldsymbol{\gamma}}')\begin{pmatrix} var\big(\hat{\boldsymbol{\xi}}'\big) & -\,var\big(\hat{\boldsymbol{\xi}}'\big) & \cdots & 0 \\ -\,var\big(\hat{\boldsymbol{\xi}}'\big) & var\big(\hat{\boldsymbol{\xi}}'\big) & \cdots & \vdots \\ \vdots & \vdots & \ddots & \vdots \\ 0 & \cdots & \cdots & 0 \end{pmatrix}\right]. \tag{C9}$$

The average across pairs takes the form

$$-trace\big[var(\hat{\boldsymbol{\gamma}}')\big(\mathbf{P} \otimes var\big(\hat{\boldsymbol{\xi}}'\big)\big)\big], \tag{C10}$$

Next, we consider (45) for a pair of genotypes. By way of analogy with (C2), we find the average of the first term across pairs to be

$$trace\big[\big(\mathbf{P} \otimes \hat{\boldsymbol{\Sigma}}_{x'}\big)\hat{\boldsymbol{\gamma}}'\hat{\boldsymbol{\gamma}}'^T\big]. \tag{C11}$$

For the second term in (45), by way of analogy with (C8), we find the average across pairs to be

$$-trace[var(\hat{\boldsymbol{\gamma}}')\,(\mathbf{P} \otimes \boldsymbol{\Sigma}_{x'})]. \tag{C12}$$

Adding up the terms in (C4), (C7), (C10), (C11), and (C12), and further adding the residual prediction variance $\bar{v}_R$, we obtain the estimated average total prediction variance of a difference given in (49).

**Appendix D**

We sourced weather covariates describing temperature, soil radiation, precipitation and air humidity from the AgERA5 database, and took an average or sum for the growing season, since we do not have information on sowing and harvest dates. This created 8 weather covariates for each dataset.

The resulting sets of covariates are listed below.

Winter rice (a total of 8):

- Precipitation sum (November-May)
- Mean solar radiation flux (November-May)
- Maximum 24-h temperature sum (November-May)
- Maximum 24-h temperature mean (November-May)
- Mean 24-h temperature (November-May)
- Minimum 24-h temperature sum (November-May)
- Minimum 24-h temperature mean (November-May)
- Relative humidity average between 6 a.m. and 6 p.m. (November-May)

Summer (a total of 8):

- Precipitation sum (June-October)
- Mean solar radiation flux (June-October)
- Maximum 24-h temperature sum (June-October)
- Maximum 24-h temperature mean (June-October)
- Mean 24-h temperature (June-October)
- Minimum 24-h temperature sum (June-October)
- Minimum 24-h temperature mean (June-October)
- Relative humidity average between 6 a.m. and 6 p.m. (June-October)

**Appendix E**

Here we provide equation for every model described in Section 4.3 for clarity. Consider model in (2), except that we factor $j$-th environment as $lm$-th location by year combination:

$y_{ilm} = \eta_{ilm} + u_{lm} + e_{ilm}$ , where $\eta_{ilm}$ is defined similar to (1): (E1)
$\eta_{ilm} = \alpha_i + \gamma_{i1} x_{lm1} + \gamma_{i2} x_{lm2} + \ldots + \gamma_{ip} x_{lmp}$ , (E2)
and contribution of $u_{lm}$ and $e_{ilm}$ was given in (46) and (47):
$u_{lm} = L_l + Y_m + (LY)_{lm}$ , (E3)
$e_{ilm} = (\alpha L)_{il} + (\alpha Y)_{im} + (\alpha LY)_{ilm}$ . (E4)
Here, $y_{ilm}$ is the observed value of $i$-th genotype in $lm$-th location by year combination (environment), $\eta_{ilm}$ pertains to the genotypic main effect $\alpha_i$ of the $i$-th genotype and genotype-specific covariate slopes $\gamma_{ik}$ for $k$-th covariate $x_{lmk}$ , $k = (1, \ldots, p)$.
Now it is important that $\gamma_{ik}$ consisst of fixed covariate slopes $\mu_{\gamma k}$ and random genotype-specific deviations $c_{ik}$:
$\gamma_{ik} = \mu_{\gamma k} + c_{ik}$ . (E5)

The models we fit differ in how $\eta_{ilm}$ and the variance-covariance of elements contained in it is specified.

**Baseline** model has:
$\gamma_{ik} = \mu_{\gamma k}$ , and thus (E6)
$y_{ilm} = \alpha_i + \mu_{\gamma 1} x_{lm1} + \ldots + \mu_{\gamma p} x_{lmp} + u_{lm} + e_{ilm}$ , (E7)
meaning only fixed covariate slopes are modelled, but not the genotype-specific deviations.

By comparison, **Kernel** approach would have $\gamma_{ik}$ as defined in (E5), but no covariance is assumed between genotype main effect and the genotype-specific deviations, so the model adheres to the form exactly as in (E1). Covariates are entering the model through a kernel matrix, so genotype-specific deviations are relating to the environmental kernel (similarity) matrix.

**RRR** of any rank would assume reduced rank variance-covariance between the genotypic intercept and genotype-specific deviations. That means, if $\gamma_{ik}$ is as in (E5), and $\boldsymbol{\mu}_\gamma^T \boldsymbol{x}_{lm} = \left(\mu_{\gamma 1} x_{lm1} + \cdots + \mu_{\gamma p} x_{lmp}\right) = \sum_{k=1}^{p} \mu_{\gamma k} x_{lmk}$ , then:
$y_{ilm} = \boldsymbol{\mu}_\gamma^T \boldsymbol{x}_{lm} + \alpha_i + c_{i1} x_{lm1} + \ldots + c_{ip} x_{lmp} + u_{lm} + e_{ilm}$ , (E8)
where $var\begin{pmatrix} a_i \\ \boldsymbol{c}_i \end{pmatrix} = \boldsymbol{\Lambda}\boldsymbol{\Lambda}^T$ , with its terms defined in (6) and (18), is approximating the unstructured variance-covariance using a set of estimated factor loadings **Λ**.

**RFR** is defined in the same way as RRR:
$y_{ilm} = \boldsymbol{\mu}_\gamma^T \boldsymbol{x}_{lm} + \alpha_i + c_{i1} x_{lm1} + \ldots + c_{ip} x_{lmp} + u_{lm} + e_{ilm}$ , (E9)

except that $var\begin{pmatrix} a_i \\ \boldsymbol{c}_i \end{pmatrix} = \boldsymbol{\Sigma}$ and is assumed to be completely unstructured.

**FW-US** model of any rank will be fitted in two steps. Step one includes extracting factor loadings in $\boldsymbol{\Lambda}_\gamma$ pertaining to the genotype-specific slopes, so that $var(\boldsymbol{c}_i) = \boldsymbol{\Lambda}_\gamma \boldsymbol{\Lambda}_\gamma^T$. The model in the first step considers genotype main effect and $u_{lm}$ as fixed and has no fixed covariate slopes:

$$y_{ilm} = \alpha_i + u_{lm} + c_{i1}x_{lm1} + \ldots + c_{ip}x_{lmp} + e_{ilm} \ . \qquad \text{(E10)}$$

These $\boldsymbol{\Lambda}_\gamma$ factor loadings are then used to construct synthetic covariates $\mathbf{Z} = \{z_{jh}\} = \mathbf{X}\boldsymbol{\Lambda}_\gamma$ , and synthetic covariates are used in the second step:

$$y_{ilm} = \boldsymbol{\mu}_\gamma^T \boldsymbol{z}_{lm} + \alpha_i \ + \ c_{i1}z_{lm1} + \ldots + c_{ih}z_{lmq} + u_{lm} + e_{ilm} \ , \qquad \text{(E11)}$$

where $var\begin{pmatrix} a_i \\ \boldsymbol{c}_i \end{pmatrix} = \boldsymbol{\Sigma}$ is assumed to be completely unstructured, and synthetic covariates $z_{lmh}$ play the same role as observed covariates, except their number is restricted to $h = (1, \ldots, q)$, that is the rank of the reduced rank variance-covariance structure in the first step.

In models that are fitted without the main EC effect $\boldsymbol{\mu}_\gamma^T \boldsymbol{x}_{lm}$ was omitted.

We can cast these models into code-like notation (e.g. based on ASReml-R 4.2), where fixed and random parts are specified separately, ":" denotes an interaction, and "$rr()$" or "$us()$" are operators that impose reduced rank or unstructured variance-covariance on selected components.

**Baseline**:

fixed = $x_1 + \ldots + x_p$ ;

random $= \ \alpha \ + L + Y + L{:}\,Y + \alpha L + \alpha Y + \alpha LY$ .

**Kinship**:

fixed = $x_1 + \ldots + x_p$ ;

random $= \ \alpha \ + \alpha{:}\,\mathbf{K} \ + L + Y + L{:}\,Y + \alpha L + \alpha Y + \alpha LY$ ,

where $\mathbf{K} = \mathbf{X}\mathbf{X}^T$ is the environmental kernel matrix.

**RRR**:

fixed = $x_1 + \ldots + x_p$ ;

random $= \ rr(\alpha \ + \alpha{:}\,x_1 + \ldots + \alpha{:}\,x_p) \ + L + Y + L{:}\,Y + \alpha L + \alpha Y + \alpha LY$ .

**RFR**:

fixed = $x_1 + \ldots + x_p$ ;

random $= \ us(\alpha \ + \alpha{:}\,x_1 + \ldots + \alpha{:}\,x_p) \ + L + Y + L{:}\,Y + \alpha L + \alpha Y + \alpha LY$ .

**FW-US**:

(step 1)

fixed = $\alpha \ + L + Y + L{:}\,Y$ ;

random $= \ rr(\alpha{:}\,x_1 + \ldots + \alpha{:}\,x_p) \ + \alpha L + \alpha Y + \alpha LY$ .

(step 2)

fixed = $z_1 + \ldots + z_h$ ;

random $= \ us(\alpha \ + \alpha{:}\,z_1 + \ldots + \alpha{:}\,z_h) \ + L + Y + L{:}\,Y + \alpha L + \alpha Y + \alpha LY$ .

In models that are fitted without the main EC effect $x_1 + \ldots + x_p$ or $z_1 + \ldots + z_h$ in the fixed part were omitted.

**Regression approaches for modelling genotype-environment interaction and making predictions into unseen environments**

Maksym Hrachov[1], Hans-Peter Piepho[1], Niaz Md. Farhat Rahman[2], Waqas Ahmed Malik[1]
1 Biostatistics Unit, Institute of Crop Science, University of Hohenheim, Stuttgart, Germany
2 Bangladesh Rice Research Institute (BRRI), Gazipur, Bangladesh

**Correspondence**
Hans-Peter Piepho, Biostatistics Unit, Institute of Crop Science, University of Hohenheim, 70593 Stuttgart, Germany.
Email: piepho@uni-hohenheim.de

**SUPPLEMENTARY INFORMATION**

**Table S1.** Model fit (winter rice) featuring the total number of parameters in the model, including those involved in synthetic covariates, full log-likelihood, and Akaike Information Criterion (AIC). Displayed values are rounded. Baseline – model without genotype-covariate interactions, Kernel – model with an environmental kernel matrix, RRR1 and RRR2 – reduced rank regression of rank one and two with observed covariates, FW1-US and FW2-US – random factorial regression with one and two synthetic covariates respectively, RFR – random factorial regression with observed covariates.

| With the main EC effect | | | | Without the main EC effect | | | |
|---|---|---|---|---|---|---|---|
| **Model** | **Parameters** | **LogLik** | **AIC** | **Model** | **Parameters** | **LogLik** | **AIC** |
| Baseline | 16 | -542 | 1117 | Baseline | 8 | -553 | 1122 |
| Kernel | 17 | -542 | 1119 | Kernel | 9 | -553 | 1124 |
| RRR1 | 24 | -525 | 1098 | RRR1 | 16 | -536 | 1104 |
| RRR2 | 32 | -505 | 1074 | RRR2 | 24 | -516 | 1079 |
| RFR | 60 | -496 | 1113 | RFR | 52 | -507 | 1118 |
| FW1-US | 19 | -519 | 1077 | FW1-US | 18 | -524 | 1084 |
| FW2-US | 30 | -507 | 1074 | FW2-US | 28 | -513 | 1082 |

**Table S2.** Variance components of models (summer rice). Baseline – model without genotype-covariate interactions, Kernel – model with an environmental kernel matrix, RRR1 and RRR2 – reduced rank regression of rank one and two with observed covariates, FW1-US and FW2-US – random factorial regression with one and two synthetic covariates respectively, RFR – random factorial regression with observed covariates, $L$ – location, $Y$ – year, $\alpha$ – genotype.

| **Models with the main EC effect** | | | | | | | |
|---|---|---|---|---|---|---|---|
| **Component** | **Baseline** | **Kernel** | **RRR1** | **RRR2** | **RFR** | **FW1-US** | **FW2-US** |
| $L$ | 0.0568 | 0.0554 | 0.0557 | 0.0536 | 0.0537 | 0.1016 | 0.0535 |
| $Y$ | 0.0262 | 0.0270 | 0.0257 | 0.0287 | 0.0287 | 0.0237 | 0.0370 |
| $\alpha$ | 0.2422 | 0.2438 | 0.2309 | 0.2321 | 0.2291 | 0.2339 | 0.2345 |
| $LY$ | 0.4434 | 0.4442 | 0.4446 | 0.4466 | 0.4462 | 0.4476 | 0.4429 |
| $\alpha L$ | 0.0307 | 0.0225 | 0.0268 | 0.0223 | 0.0186 | 0.0249 | 0.0218 |
| $\alpha Y$ | 0.0142 | 0.0108 | 0.0120 | 0.0071 | 0.0060 | 0.0097 | 0.0096 |
| $\alpha LY$ | 0.2695 | 0.2693 | 0.2706 | 0.2678 | 0.2668 | 0.2667 | 0.2666 |
| **Models without the main EC effect** | | | | | | | |
| **Component** | **Baseline** | **Kernel** | **RRR1** | **RRR2** | **RFR** | **FW1-US** | **FW2-US** |
| $L$ | 0.1269 | 0.1275 | 0.1295 | 0.1301 | 0.1308 | 0.1280 | 0.1299 |
| $Y$ | 0.0250 | 0.0263 | 0.0274 | 0.0321 | 0.0320 | 0.0300 | 0.0297 |
| $\alpha$ | 0.2430 | 0.2450 | 0.2328 | 0.2344 | 0.2316 | 0.2352 | 0.2359 |
| $LY$ | 0.4441 | 0.4448 | 0.4459 | 0.4475 | 0.4473 | 0.4459 | 0.4472 |
| $\alpha L$ | 0.0307 | 0.0226 | 0.0269 | 0.0221 | 0.0186 | 0.0248 | 0.0218 |
| $\alpha Y$ | 0.0142 | 0.0108 | 0.0119 | 0.0071 | 0.0060 | 0.0097 | 0.0095 |
| $\alpha LY$ | 0.2695 | 0.2693 | 0.2706 | 0.2678 | 0.2669 | 0.2668 | 0.2667 |

**Table S3.** Variance components of models (winter rice). Baseline – model without genotype-covariate interactions, Kernel – model with an environmental kernel matrix, RRR1 and RRR2 – reduced rank regression of rank one and two with observed covariates, FW1-US and FW2-US – random factorial regression with one and two synthetic covariates respectively, RFR – random factorial regression with observed covariates, $L$ – location, $Y$ – year, $\alpha$ – genotype.

| **Models with the main EC effect** | | | | | | | |
|---|---|---|---|---|---|---|---|
| **Component** | **Baseline** | **Kernel** | **RRR1** | **RRR2** | **FW1-US** | **FW2-US** | **RFR** |
| $L$ | 0.1530 | 0.1530 | 0.1531 | 0.1537 | 0.1529 | 0.1987 | 0.1776 |
| $Y$ | 0.0381 | 0.0381 | 0.0383 | 0.0391 | 0.0388 | 0.0361 | 0.0301 |
| $\alpha$ | 0.1331 | 0.1331 | 0.1034 | 0.1119 | 0.1152 | 0.1115 | 0.1104 |
| $LY$ | 0.4932 | 0.4932 | 0.4928 | 0.4930 | 0.4934 | 0.5030 | 0.5077 |
| $\alpha L$ | 0.0310 | 0.0310 | 0.0304 | 0.0308 | 0.0274 | 0.0321 | 0.0304 |
| $\alpha Y$ | 0.0148 | 0.0148 | 0.0134 | 0.0058 | 0.0041 | 0.0066 | 0.0051 |
| $\alpha LY$ | 0.2593 | 0.2593 | 0.2583 | 0.2584 | 0.2569 | 0.2591 | 0.2573 |
| **Models without the main EC effect** | | | | | | | |
| $L$ | 0.2126 | 0.2126 | 0.2160 | 0.2156 | 0.2135 | 0.2129 | 0.2155 |
| $Y$ | 0.0435 | 0.0435 | 0.0468 | 0.0433 | 0.0425 | 0.0443 | 0.0445 |
| $\alpha$ | 0.1338 | 0.1338 | 0.1051 | 0.1154 | 0.1185 | 0.1146 | 0.1135 |
| $LY$ | 0.5198 | 0.5198 | 0.5210 | 0.5184 | 0.5187 | 0.5202 | 0.5208 |
| $\alpha L$ | 0.0310 | 0.0309 | 0.0303 | 0.0309 | 0.0274 | 0.0322 | 0.0305 |
| $\alpha Y$ | 0.0148 | 0.0147 | 0.0134 | 0.0057 | 0.0040 | 0.0064 | 0.0050 |
| $\alpha LY$ | 0.2593 | 0.2593 | 0.2583 | 0.2584 | 0.2569 | 0.2591 | 0.2573 |

**Table S4.** Variance components present change relative to the baseline (winter rice). Baseline – model without genotype-covariate interactions, Kernel – model with an environmental kernel matrix, RRR1 and RRR2 – reduced rank regression of rank one and two with observed covariates, FW1-US and FW2-US – random factorial regression with one and two synthetic covariates respectively, RFR – random factorial regression with observed covariates, $L$ – location, $Y$ – year, $\alpha$ – genotype.

| **Models with the main EC effect** | | | | | | | |
|---|---|---|---|---|---|---|---|
| **Component** | **Baseline**[a] | **Kernel** | **RRR1** | **RRR2** | **RFR** | **FW1-US** | **FW2-US** |
| $L$ | 0.1530 | 0.0 | 0.1 | 0.5 | -0.1 | 29.9 | 16.1 |
| $Y$ | 0.0381 | 0.0 | 0.5 | 2.7 | 1.9 | -5.2 | -20.9 |
| $\alpha$ | 0.1331 | 0.0 | -22.3 | -15.9 | -13.4 | -16.3 | -17.0 |
| $LY$ | 0.4932 | 0.0 | -0.1 | 0.0 | 0.0 | 2.0 | 2.9 |
| $\alpha L$ | 0.0310 | -0.1 | -1.9 | -0.6 | -11.7 | 3.5 | -1.9 |
| $\alpha Y$ | 0.0148 | -0.1 | -9.3 | -60.5 | -72.0 | -55.5 | -65.5 |
| $\alpha LY$ | 0.2593 | 0.0 | -0.4 | -0.3 | -0.9 | -0.1 | -0.8 |
| **Models without the main EC effect** | | | | | | | |
| **Component** | **Baseline**[a] | **Kernel** | **RRR1** | **RRR2** | **RFR** | **FW1-US** | **FW2-US** |
| $L$ | 0.2126 | 0.0 | 1.6 | 1.4 | 0.4 | 0.1 | 1.4 |
| $Y$ | 0.0435 | 0.0 | 7.7 | -0.4 | -2.2 | 1.9 | 2.3 |
| $\alpha$ | 0.1338 | 0.0 | -21.5 | -13.8 | -11.5 | -14.4 | -15.2 |
| $LY$ | 0.5198 | 0.0 | 0.2 | -0.3 | -0.2 | 0.1 | 0.2 |
| $\alpha L$ | 0.0310 | -0.1 | -2.1 | -0.2 | -11.4 | 3.9 | -1.5 |
| $\alpha Y$ | 0.0148 | -0.1 | -9.3 | -61.6 | -72.8 | -56.3 | -66.4 |
| $\alpha LY$ | 0.2593 | 0.0 | -0.4 | -0.5 | -1.2 | -0.2 | -0.8 |

[a] Baseline represents the actual values from the Baseline model, the rest are in percent relative to it.

**Table S5.** Leave-one-environment-out (LOEO) and leave-one-year-and-location-out (LYLO) cross-validation means (winter rice). PCC – Pearson's correlation coefficient, MSEPD – mean squared error of prediction difference, MSPE – mean squared prediction error, Baseline – model without genotype-covariate interactions, Kernel – model with an environmental kernel matrix, RRR1 and RRR2 – reduced rank regression of rank one and two with observed covariates, FW1-US and FW2-US – random factorial regression with one and two synthetic covariates, RFR – random factorial regression with observed covariates.

| Type | Model | Mean PCC | | Mean MSEPD | | Mean MSPE | |
|---|---|---|---|---|---|---|---|
| | | LOEO | LYLO | LOEO | LYLO | LOEO | LYLO |
| With the main EC effect | Baseline | 0.466 | 0.400 | 0.780 | 0.838 | 0.960 | 1.22 |
| | Kernel | 0.466 | 0.400 | 0.780 | 0.838 | 0.960 | 1.21 |
| | RRR1 | 0.468 | 0.393 | 0.780 | 0.846 | 0.960 | 1.22 |
| | RRR2 | 0.465 | 0.389 | 0.779 | 0.850 | 0.962 | 1.22 |
| | RFR | 0.462 | 0.389 | 0.782 | 0.849 | 0.964 | 1.22 |
| | FW1-US | 0.468 | 0.391 | 0.777 | 0.847 | 0.957 | 1.22 |
| | FW2-US | 0.465 | 0.389 | 0.780 | 0.85 | 0.962 | 1.23 |
| Without the main EC effect | Baseline | 0.466 | 0.400 | 0.780 | 0.838 | 0.969 | 1.22 |
| | Kernel | 0.466 | 0.400 | 0.780 | 0.838 | 0.969 | 1.22 |
| | RRR1 | 0.468 | 0.393 | 0.780 | 0.846 | 0.971 | 1.22 |
| | RRR2 | 0.465 | 0.389 | 0.779 | 0.850 | 0.968 | 1.22 |
| | RFR | 0.462 | 0.389 | 0.782 | 0.849 | 0.970 | 1.22 |
| | FW1-US | 0.467 | 0.391 | 0.777 | 0.847 | 0.966 | 1.22 |
| | FW2-US | 0.465 | 0.388 | 0.780 | 0.850 | 0.970 | 1.22 |

**Table S6.** Leave-one-environment-out (LOEO) and leave-one-year-and-location-out (LYLO) cross-validation medians (winter rice). PCC – Pearson's correlation coefficient, MSEPD – mean squared error of prediction difference, MSPE – mean squared prediction error, Baseline – model without genotype-covariate interactions, Kernel – model with an environmental kernel matrix, RRR1 and RRR2 – reduced rank regression of rank one and two with observed covariates, FW1-US and FW2-US – random factorial regression with one and two synthetic covariates, RFR – random factorial regression with observed covariates.

| Type | Model | Median PCC | | Median MSEPD | | Median MSPE | |
|---|---|---|---|---|---|---|---|
| | | **LOEO** | **LYLO** | **LOEO** | **LYLO** | **LOEO** | **LYLO** |
| With the main EC effect | Baseline | 0.545 | 0.438 | 0.661 | 0.711 | 0.620 | 0.744 |
| | Kernel | 0.545 | 0.435 | 0.661 | 0.712 | 0.620 | 0.797 |
| | RRR1 | 0.544 | 0.430 | 0.658 | 0.701 | 0.622 | 0.816 |
| | RRR2 | 0.547 | 0.433 | 0.664 | 0.723 | 0.653 | 0.810 |
| | RFR | 0.553 | 0.427 | 0.678 | 0.722 | 0.634 | 0.804 |
| | FW1-US | 0.546 | 0.442 | 0.646 | 0.722 | 0.631 | 0.863 |
| | FW2-US | 0.540 | 0.430 | 0.692 | 0.731 | 0.643 | 0.851 |
| Without the main EC effect | Baseline | 0.546 | 0.437 | 0.661 | 0.711 | 0.658 | 0.824 |
| | Kernel | 0.546 | 0.435 | 0.661 | 0.712 | 0.658 | 0.824 |
| | RRR1 | 0.544 | 0.431 | 0.657 | 0.702 | 0.664 | 0.851 |
| | RRR2 | 0.547 | 0.432 | 0.665 | 0.725 | 0.633 | 0.851 |
| | RFR | 0.553 | 0.427 | 0.677 | 0.723 | 0.630 | 0.839 |
| | FW1-US | 0.544 | 0.443 | 0.647 | 0.723 | 0.642 | 0.846 |
| | FW2-US | 0.539 | 0.431 | 0.693 | 0.734 | 0.634 | 0.846 |

**Table S7.** Leave-one-environment-out (LOEO) and leave-one-year-and-location-out (LYLO) cross-validation medians (summer rice). PCC – Pearson's correlation coefficient, MSEPD – mean squared error of prediction difference, MSPE – mean squared prediction error, Baseline – model without genotype-covariate interactions, Kernel – model with an environmental kernel matrix, RRR1 and RRR2 – reduced rank regression of rank one and two with observed covariates, FW1-US and FW2-US – random factorial regression with one and two synthetic covariates, RFR – random factorial regression with observed covariates.

| **Type** | **Model** | **Median PCC** | | **Median MSEPD** | | **Median MSPE** | |
|---|---|---|---|---|---|---|---|
| | | **LOEO** | **LYLO** | **LOEO** | **LYLO** | **LOEO** | **LYLO** |
| With the main EC effect | Baseline | 0.697 | 0.665 | 0.645 | 0.699 | 0.576 | 0.656 |
| | Kernel | 0.695 | 0.671 | 0.617 | 0.697 | 0.570 | 0.645 |
| | RRR1 | 0.695 | 0.668 | 0.665 | 0.709 | 0.569 | 0.660 |
| | RRR2 | 0.691 | 0.671 | 0.654 | 0.720 | 0.581 | 0.665 |
| | RFR | 0.693 | 0.661 | 0.652 | 0.720 | 0.575 | 0.665 |
| | FW1-US | 0.694 | 0.669 | 0.648 | 0.717 | 0.556 | 0.642 |
| | FW2-US | 0.694 | 0.669 | 0.665 | 0.716 | 0.605 | 0.625 |
| Without the main EC effect | Baseline | 0.697 | 0.665 | 0.645 | 0.699 | 0.584 | 0.644 |
| | Kernel | 0.695 | 0.670 | 0.618 | 0.697 | 0.582 | 0.652 |
| | RRR1 | 0.695 | 0.668 | 0.665 | 0.711 | 0.582 | 0.650 |
| | RRR2 | 0.691 | 0.670 | 0.654 | 0.720 | 0.567 | 0.672 |
| | RFR | 0.693 | 0.662 | 0.653 | 0.720 | 0.561 | 0.678 |
| | FW1-US | 0.694 | 0.668 | 0.648 | 0.716 | 0.571 | 0.671 |
| | FW2-US | 0.694 | 0.668 | 0.665 | 0.716 | 0.572 | 0.672 |

**Table S8.** Mean variance of prediction (MVP) and mean squared prediction error (MSPE) from leave-one-year-and-location-out cross-validation (winter rice). Baseline – model without genotype-covariate interactions, Kernel – model with an environmental kernel matrix, RRR1 and RRR2 – reduced rank regression of rank one and two with observed covariates, FW1-US and FW2-US – random factorial regression with one and two synthetic covariates, RFR – random factorial regression with observed covariates. MVP for the baseline model was taken from the standard ASReml-R output.

| Type | Model | MSPE | | MVP | |
|---|---|---|---|---|---|
| | | Mean | Median | Mean | Median |
| With the main EC effect | Baseline | 1.22 | 0.744 | 1.10 | 1.10 |
| | RRR1 | 1.22 | 0.816 | 1.17 | 1.17 |
| | RRR2 | 1.22 | 0.810 | 1.17 | 1.17 |
| | RFR | 1.22 | 0.804 | 1.16 | 1.16 |
| | FW1-US | 1.22 | 0.863 | 1.11 | 1.12 |
| | FW2-US | 1.23 | 0.851 | 1.08 | 1.08 |
| Without the main EC effect | Baseline | 1.22 | 0.824 | 1.12 | 1.12 |
| | RRR1 | 1.22 | 0.851 | 1.13 | 1.13 |
| | RRR2 | 1.22 | 0.851 | 1.12 | 1.13 |
| | RFR | 1.22 | 0.839 | 1.12 | 1.12 |
| | FW1-US | 1.22 | 0.846 | 1.12 | 1.12 |
| | FW2-US | 1.22 | 0.846 | 1.12 | 1.13 |

**Table S9.** Variance of the predicted difference (VPD) and mean squared error of prediction difference (MSEPD) from the leave-one-year-and-location-out cross-validation (winter rice). Baseline – model without genotype-covariate interactions, Kernel – model with an environmental kernel matrix, RRR1 and RRR2 – reduced rank regression of rank one and two with observed covariates, FW1-US and FW2-US – random factorial regression with one and two synthetic covariates, RFR – random factorial regression with observed covariates.

| Type | Model | MSEPD | | VPD | |
|---|---|---|---|---|---|
| | | **Mean** | **Median** | **Mean** | **Median** |
| With the main EC effect | Baseline | 0.838 | 0.711 | 0.631 | 0.631 |
| | RRR1 | 0.846 | 0.701 | 0.626 | 0.634 |
| | RRR2 | 0.850 | 0.723 | 0.612 | 0.628 |
| | RFR | 0.849 | 0.722 | 0.624 | 0.614 |
| | FW1-US | 0.847 | 0.722 | 0.614 | 0.624 |
| | FW2-US | 0.85 | 0.731 | 0.635 | 0.616 |
| Without the main EC effect | Baseline | 0.838 | 0.711 | 0.631 | 0.631 |
| | RRR1 | 0.846 | 0.702 | 0.635 | 0.634 |
| | RRR2 | 0.850 | 0.725 | 0.626 | 0.628 |
| | RFR | 0.849 | 0.723 | 0.612 | 0.614 |
| | FW1-US | 0.847 | 0.723 | 0.624 | 0.624 |
| | FW2-US | 0.85 | 0.734 | 0.614 | 0.615 |

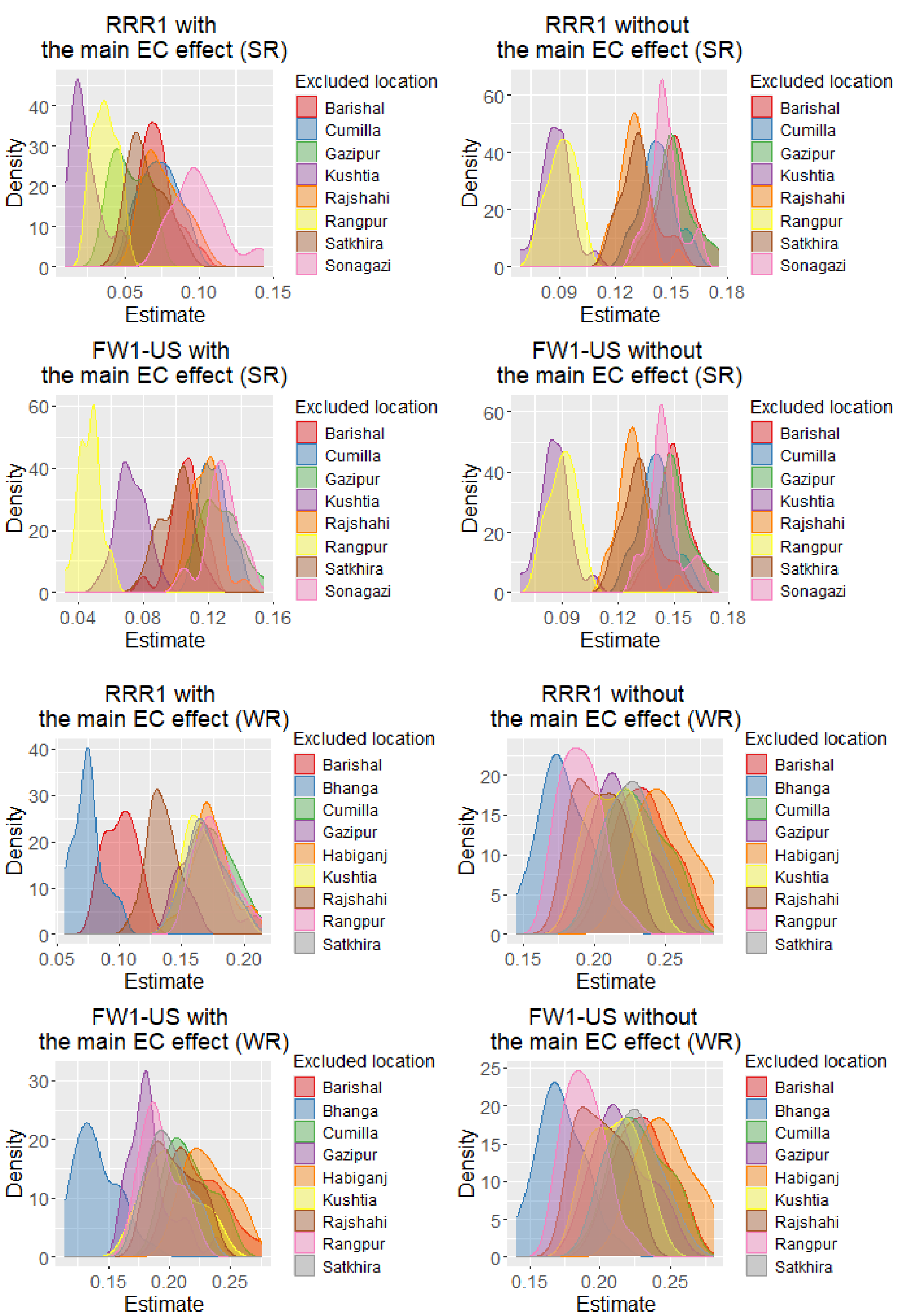

**Figure S1.** The distribution of the location (*L*) variance component estimate in leave-one-year-and-location-out (LYLO) cross-validation (CV) for the summer rice (SR) and the winter rice (WR) datasets. RRR1 and RRR2 – reduced rank regression of rank one and two with observed covariates, FW1-US and FW2-US – random factorial regression with one and two synthetic covariates.

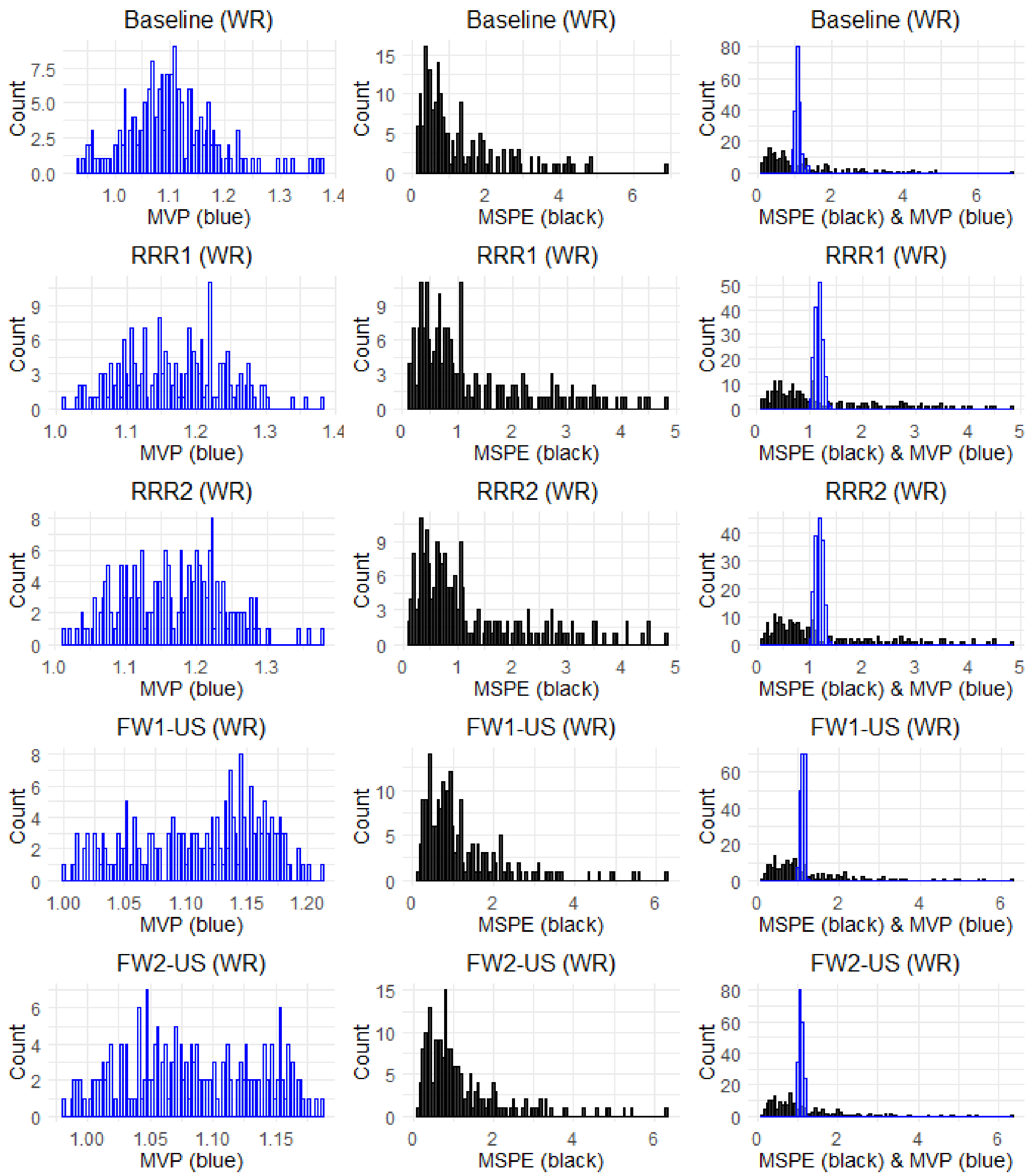

**Figure S2.** Distributions of model-based mean variance of prediction and cross-validation-based mean squared prediction error for all models with the main EC effect fitted to the winter rice (WR) data. Baseline – model without genotype-covariate interactions, RRR1 and RRR2 – reduced rank regression of rank one and two with observed covariates, FW1-US and FW2-US – random factorial regression with one and two synthetic covariates.

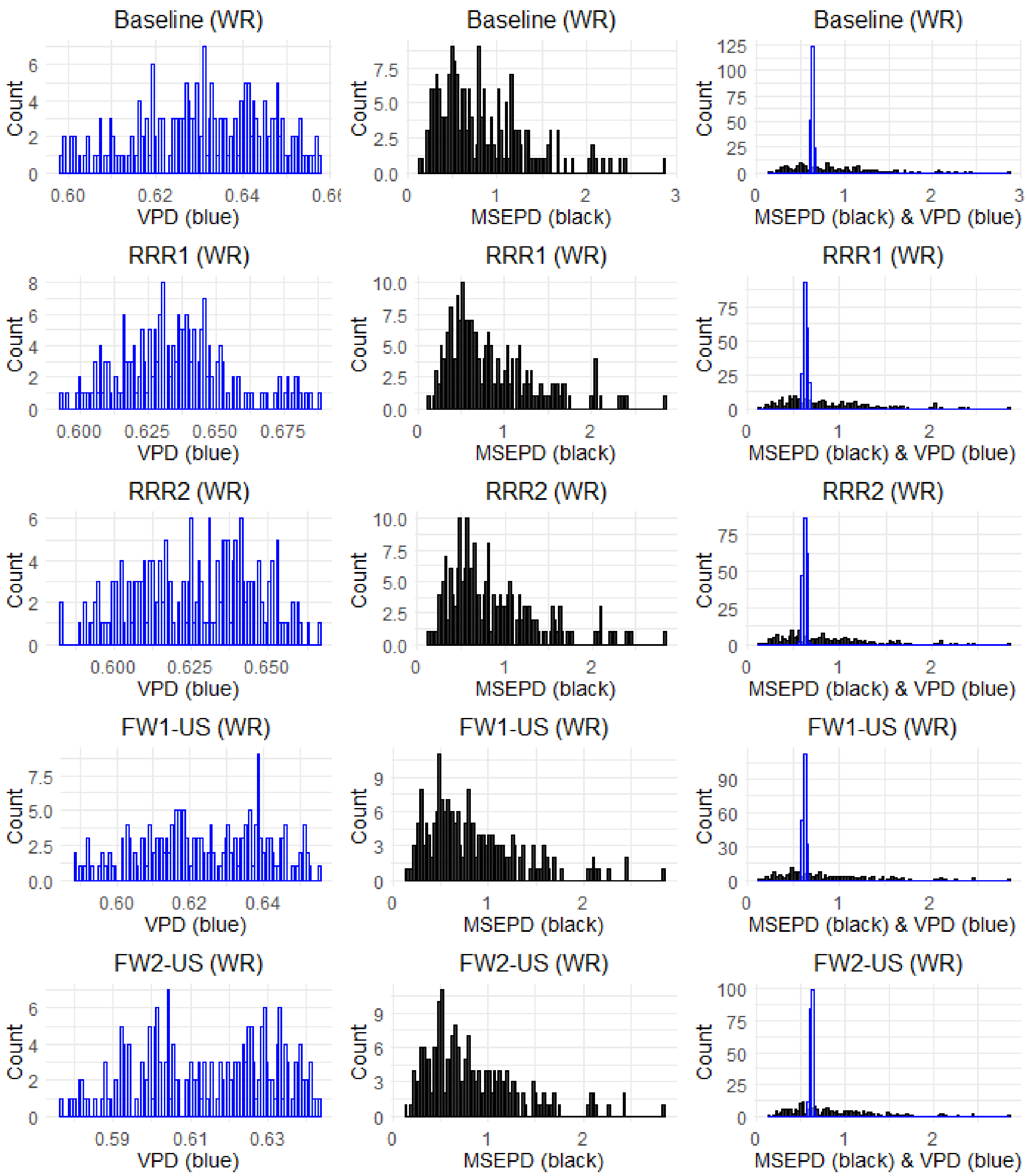

**Figure S3.** Distributions of model-based variance of predicted difference and cross-validation-based mean squared error of prediction difference for all models with the main EC effect fitted to the winter rice (WR) data. Baseline – model without genotype-covariate interactions, RRR1 and RRR2 – reduced rank regression of rank one and two with observed covariates, FW1-US and FW2-US – random factorial regression with one and two synthetic covariates.

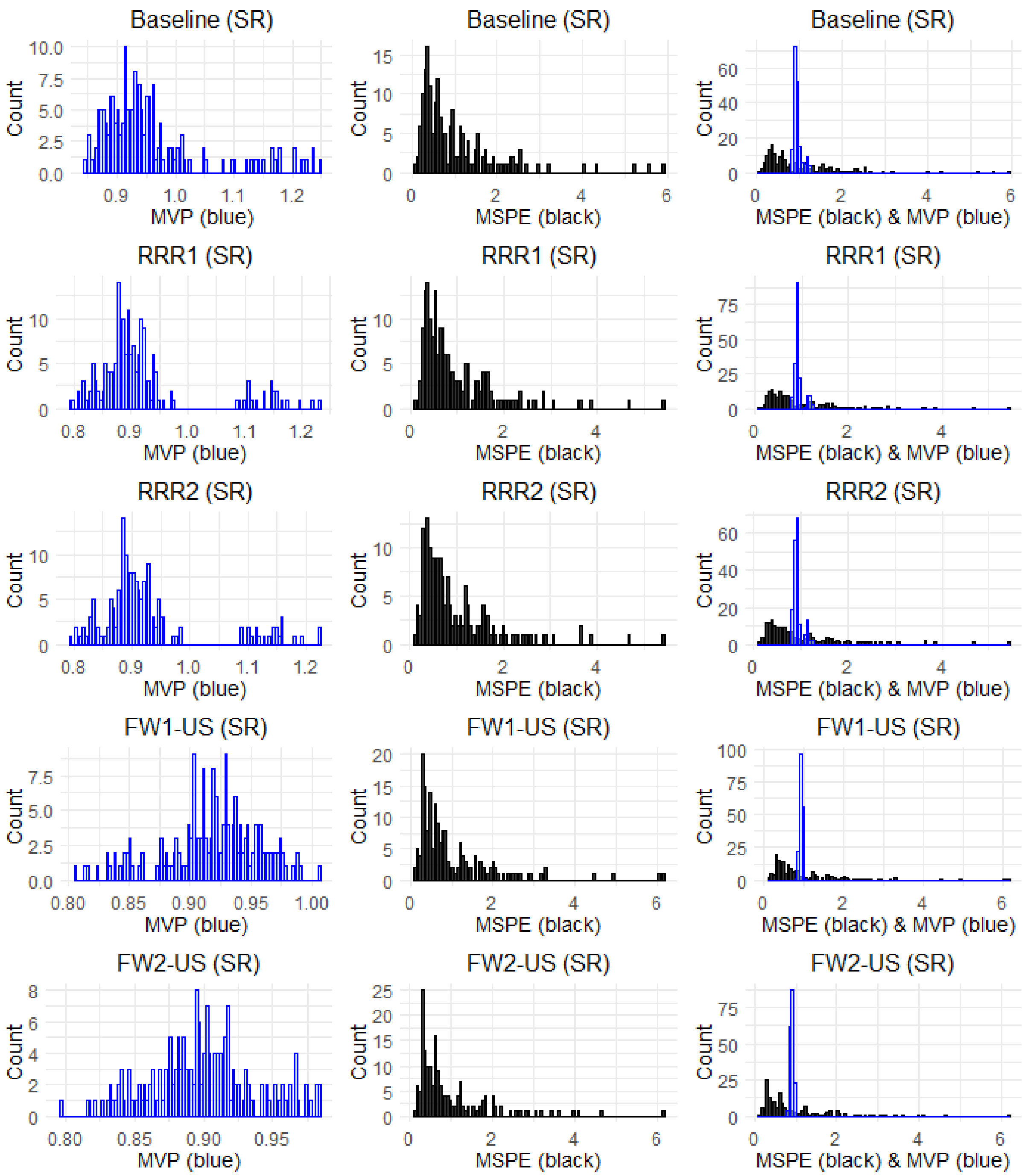

**Figure S4.** Distributions of model-based mean variance of prediction and cross-validation-based mean squared prediction error for all models with the main EC effect fitted to the summer rice (SR) data. Baseline – model without genotype-covariate interactions, RRR1 and RRR2 – reduced rank regression of rank one and two with observed covariates, FW1-US and FW2-US – random factorial regression with one and two synthetic covariates.

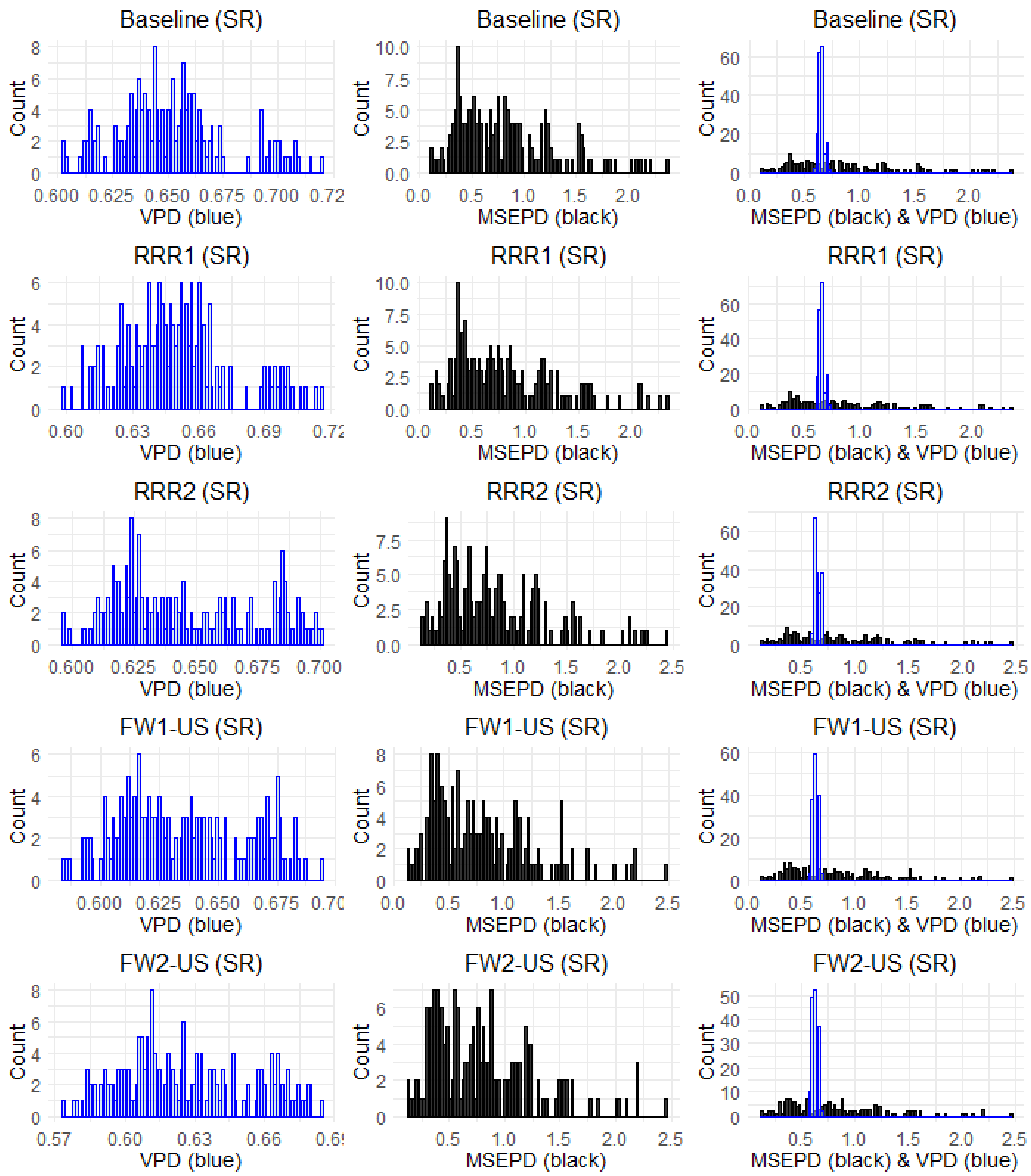

**Figure S5.** Distributions of model-based variance of predicted difference and cross-validation-based mean squared error of prediction difference for all models with the main EC effect fitted to the summer rice (SR) data. Baseline – model without genotype-covariate interactions, RRR1 and RRR2 – reduced rank regression of rank one and two with observed covariates, FW1-US and FW2-US – random factorial regression with one and two synthetic covariates.

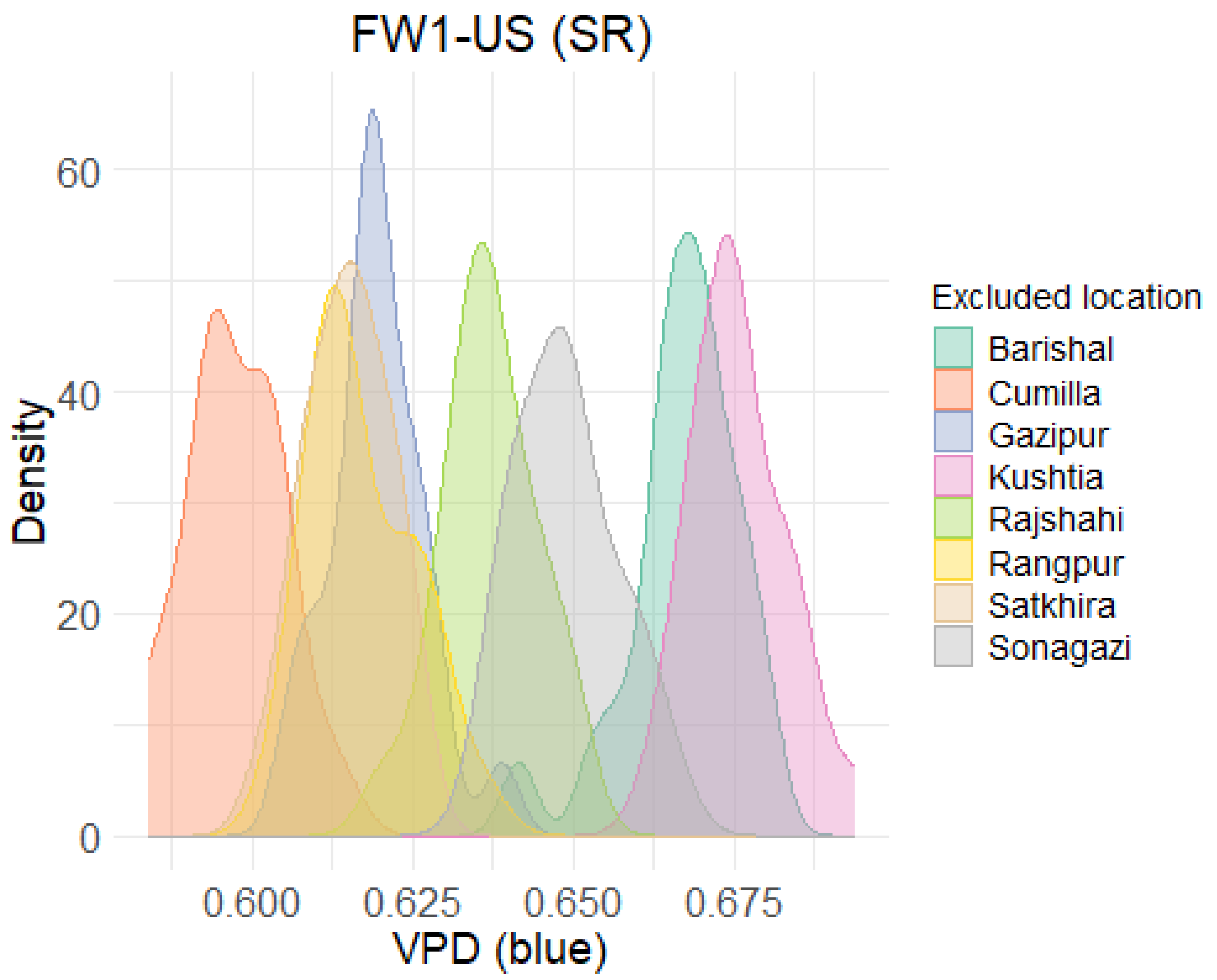

**Figure S6.** The bimodal distribution observed in Fig. S5 is due to exclusion of locations in the leave-one-year-and-location-out (LYLO) cross-validation approach. Different colors represent which location was excluded. VPD – variance of the predicted difference, FW1-US – random factorial regression with one synthetic covariate (with the main EC effect).